\documentstyle[12pt,epsfig]{article}
\def\lbldef#1#2{\expandafter\gdef\csname #1\endcsname {#2}}

\def\href#1#2{#2}  

\begin{document}
\baselineskip=15.5pt
\pagestyle{plain}
\setcounter{page}{1}

\begin{titlepage}

\begin{flushright}
CERN-TH-99-264\\
hep-th/9908182
\end{flushright}
\vspace{10 mm}

\begin{center}
{\Large  Delocalized Supergravity Solutions for Brane/Anti-brane Systems 
and their Bound States}

\vspace{5mm}

\end{center}

\vspace{5 mm}

\begin{center}
{\large Donam Youm\footnote{Donam.Youm@cern.ch}}

\vspace{3mm}

Theory Division, CERN, CH-1211, Geneva 23, Switzerland

\end{center}

\vspace{1cm}

\begin{center}
{\large Abstract}
\end{center}

\noindent

We obtain various solutions for $D=4$ dipoles and their bound states 
whose $U(1)$ fields originate from various form fields in the effective 
string theories.  We oxidize such dipole solutions to $D=10$ to obtain 
delocalized supergravity solutions for the brane/anti-brane pairs and 
their bound states. We speculate on generalized harmonic superposition 
rules for supergravity solutions for (intersecting) brane/anti-brane 
pairs. 

\vspace{1cm}
\begin{flushleft}
CERN-TH-99-264\\
August, 1999
\end{flushleft}
\end{titlepage}
\newpage

\section{Introduction}

Due to the recent development in string dualities, our understanding 
of string theories has greatly advanced.  Initially, the duality relations 
between different string theories are established by relating the effective 
supergravity actions (namely the massless string vibration modes) 
of string theories through field redefinitions.  Such duality relations are 
further extended to the (massive) BPS states.  The BPS states are useful in 
establishing and proving the duality relations because of their stability 
and the absence of quantum corrections due to the supersymmetry preserved 
by such states.  So, their properties can be safely extrapolated from weak 
string coupling to strong string coupling, thereby allowing test of 
non-perturbative string dualities, i.e. $S$-dualities and $U$-dualities, 
and the study of non-perturbative aspects of string theories.

If string dualities are correct, one should be able to establish duality 
relations beyond the massless and the BPS states.  The pioneering work 
\cite{sen1} by Sen first established the duality relations among non-BPS 
states.  In perturbative spectrum of a string theory, there are some stable 
(due to their being the {\it lightest} and therefore unable to decay to 
other states) non-BPS states , i.e. there are no particular relations 
between their masses and their charges, with some conserved quantum numbers.   
Since such states do not preserve supersymmetry, they receive quantum 
corrections.  Nonetheless, due to their stability such states should be 
present also in the strong coupling region as non-perturbative states.  
The stable non-BPS perturbative string states considered in Ref. \cite{sen1} 
are the lightest massive perturbative states in the $SO(32)$ heterotic string 
belonging to the spinor representation of $SO(32)$.  These states are not 
BPS since the theory that such states belong to does not have the central 
charge.  If the duality between the $SO(32)$ heterotic string and the type 
I theory \cite{wit1,dab,hul,pwit} is correct, then there should exist the 
corresponding stable non-BPS states with the same quantum numbers within 
the non-perturbative spectrum of the type I string theory.  Such stable 
non-BPS state with the same quantum numbers in the type I string theory 
can be identified \cite{pol} as a tachyonic kink solution on the 
D-string/anti-D-string pair. Such solution is shown \cite{sen2} to be 
stable and behave as a D-particle.  Generally, D-brane/anti-D-brane systems 
are useful for studying non-perturbative aspects of string theories beyond 
the BPS regime.  

Therefore, it is interesting to construct supergravity solutions for 
brane/anti-brane pairs.  The first brane/anti-brane pair solution that 
was constructed is the D6-brane/anti-D6-brane pair solution in Ref. 
\cite{sen3}.  Sen constructed such solution by embedding the magnetic 
Kaluza-Klein (KK) dipole solutions of Ref. \cite{gper} into $D=11$ 
supergravity and then compactifying on $S^1$.  Such solution is a static 
and stable configuration because a net attractive force between the brane 
and the anti-brane due to the gravitational and the electromagnetic 
interactions is cancelled by the repulsive force between the brane and the 
anti-brane induced by the external magnetic field.  For supergravity 
solutions for the other brane/anti-brane systems, it seems to be a 
challenging problem at this moment to construct them.  Unlike the brane 
solutions, which can be straightforwardly constructed by the solution 
generating transformations, which induce brane charges on a charge neutral 
solution, currently there is not available systematic method to construct 
solutions for general brane/anti-brane pairs.  Even the generalization of 
the Bonnor-transformation \cite{bon}, which was used to construct dipole 
solutions from the $D=4$ Kerr solution, and the method used in Ref. 
\cite{gper}, which constructs the magnetic KK dipole solution through an 
Euclidean rotation of the $D=4$ Kerr solution and the addition of another 
time coordinate followed by the dimensional reduction, to the case of 
general brane/anti-brane systems seems to be not plausible.

In this paper, we shall construct supergravity solutions for various 
brane/anti-brane systems and their bound states by embedding $D=4$ dipole 
solutions in Einstein-Maxwell-Dilaton system into string theories.  
(Refs. \cite{muk1,muk2,emp} also study the related issues.)
Although such $D=10$ supergravity solutions have only three localized 
(overall) transverse coordinates (because these are uplifted from $D=4$ 
solutions), these might enable one to gain insight on the structure of the 
complete supergravity solutions or perhaps to guess the general Ansatz for 
the complete solutions.  So, in this paper, we present a large number of such 
supergravity solutions for the purpose of revealing the general patterns for 
the structure of the solutions.  The paper is organized as follows.  In 
section 2, we summarize general dipole solutions in the 
Einstein-Maxwell-dilaton system and then we construct dyonic dipole solution 
in the Einstein-Maxwell-dilaton-axion system by applying the $SL(2,{\bf R})$ 
transformation to the dipole solution in the Einstein-Maxwell-dilaton theory 
with the dilaton coupling $\alpha=1$.  In section 3, we discuss (consistently) 
truncated string effective action, into which the dipole solutions in the 
$D=4$ Einstein-Maxwell-Dilaton system are to be embedded, and the symmetries 
of such action to be used to generate other $D=4$ dipole solutions in 
the effective string theory.  In section 4, we present various $D=4$ 
``fundamental'' dipole solutions charged with respect to $U(1)$ fields 
originated from various form fields in string theories and then we uplift 
such solutions to $D=10$ to obtain delocalized supergravity solutions for 
various brane/anti-brane systems.  In section 5, we construct solutions 
for non-marginal bound states of ``fundamental'' $D=4$ dipoles of section 4 
by applying the transformations in section 3, and then uplift them to $D=10$ 
to obtain the delocalized supergravity solutions for the non-marginal bound 
states of brane/anti-brane systems.  In section 6, we construct delocalized 
supergravity solutions for the marginal bound states of brane/anti-brane 
systems with the equal dipole moments by embedding the $D=4$ dipole 
solutions in the Einstein-Maxwell-dilaton theory with the dilaton couplings 
$\alpha=1,1/\sqrt{3},0$.  We learn that such restricted class of supergravity 
solutions still satisfy the rules similar to the harmonic superposition 
rules of the (delocalized intersecting) brane solutions.

\section{Dipoles in Einstein-Maxwell-Dilaton System}

The gravity solution for the dipole in $D=4$ was first constructed within 
the Einstein-Maxwell system in Ref. \cite{bon} by applying the 
Bonnor-transformation to the complexified Kerr solution.  Later, the 
magnetic dipole solution in the five-dimensional KK theory was 
constructed in Ref. \cite{gper} through an Euclidean rotation of the Kerr 
solution followed by the addition of a new time coordinate and the dimensional 
reduction.  In this section, we summarize the $D=4$ dipole solutions that 
generalize these dipole solutions.  

We consider the dipole solutions in the following general 
Einstein-Maxwell-dilaton system in $D=4$:
\begin{equation}
S={1\over{2\kappa^2_4}}\int dx^4\sqrt{-g}\left[{\cal R}_g-2(\partial
\phi)^2-{1\over 4}e^{-2\alpha\phi}F^2\right],
\label{einmaxdilact}
\end{equation}
where $\phi$ is the $D=4$ dilaton and $\alpha$ is the dilaton coupling 
parameter to the $U(1)$ field strength $F$.  The cases $\alpha=0,\sqrt{3}$ 
respectively correspond to the Einstein-Maxwell system and the $D=5$ 
KK theory considered in Refs. \cite{bon} and \cite{gper}.
In particular, in the case where $\alpha=\sqrt{3},1,1/\sqrt{3},0$, 
this action can be embedded as the effective supergravity action for string 
theories \cite{dkmr}, thereby allowing the study of the solutions to the 
equations of motions for such action within the context of string theory.  

The following general dipole solution of this system was constructed in 
Refs. \cite{dged,ggk}:
\begin{eqnarray}
g_{\mu\nu}dx^{\mu}dx^{\nu}&=&-\left({{r^2-2mr-a^2\cos^2\theta}\over
{r^2-a^2\cos^2\theta}}\right)^{2\over{1+\alpha^2}}dt^2
\cr
& &+{{[(r^2-2mr-a^2\cos^2\theta)(r^2-a^2\cos^2\theta)]^{2\over{1+\alpha^2}}}
\over{(r^2-2mr+m^2\sin^2\theta-a^2\cos^2\theta)^{{3-\alpha^2}\over
{1+\alpha^2}}}}\left[{{dr^2}\over{r^2-2mr-a^2}}+d\theta^2\right]
\cr
& &+\left({{r^2-a^2\cos^2\theta}\over{r^2-2mr-a^2\cos^2\theta}}
\right)^{2\over{1+\alpha^2}}(r^2-2mr-a^2)\sin^2\theta d\varphi^2,
\cr
\phi&=&\pm{\alpha\over{1+\alpha^2}}\ln{{r^2-2mr-a^2\cos^2\theta}\over
{r^2-a^2\cos^2\theta}},
\label{dipsol}
\end{eqnarray}
where the plus [minus] sign in the dilaton solution is for the 
electric [magnetic] dipole, and the non-zero components of the $U(1)$ gauge 
potential $A_{\mu}$ for the electric and magnetic cases are respectively 
given by
\begin{eqnarray}
A_t&=&{1\over\sqrt{1+\alpha^2}}{{4ma\cos\theta}\over{r^2-a^2\cos^2\theta}},
\cr
A_{\varphi}&=&{1\over\sqrt{1+\alpha^2}}{{4mar\sin^2\theta}\over{r^2-2mr-
a^2\cos^2\theta}}.
\label{elecmagdip}
\end{eqnarray}
The ADM mass $M$ and the electric (or magnetic) dipole moment $p$ of the 
above solution are given by
\begin{equation}
M={{2m}\over{1+\alpha^2}},\ \ \ \ \ 
p={{2ma}\over\sqrt{1+\alpha^2}}.
\label{massdipol}
\end{equation}
The $\alpha=0$ and $\alpha=\sqrt{3}$ cases respectively correspond to 
the dipole solutions constructed in Ref. \cite{bon} and Ref. \cite{gper}.

The solution (\ref{dipsol}) describes a pair of dilatonic 0-brane 
\cite{hs} and anti-0-brane separated by the distance $2a$.  The endpoints 
of such dipole are located at $(r,\theta)=(r_+,0)$ and $(r_+,\pi)$, where 
$r_+=m+\sqrt{m^2+a^2}$.  Generally, the solution (\ref{dipsol}) has a 
singularity at $r=r_+$ due to the conical deficit along the line $r=r_+$.  
Furthermore, the horizon of the each hole at one end point of the 
dipole is deformed due to the field created by the other hole on the 
other end.  Therefore, in the near-region of the endpoints of the dipole, 
the metric (\ref{dipsol}) takes the form of the deformed near-horizon metric 
of the 0-brane (or anti-0-brane) due to the conical singularity.  
Nevertheless, the solution (\ref{dipsol}) can still be interpreted as pair 
of 0-brane and anti-0-brane kept separated by struts or strings 
\cite{dged,muk1}.  In the case of the magnetic KK dipole solution of Ref. 
\cite{gper}, the conical singularity at $r=r_+$ is removed by identifying 
points under a combined spatial translation (along the internal coordinate 
direction) and rotation.  In fact, such transformation on the 
Minkowski spacetime (in cylindrical coordinates) leads to the space 
with an infinitely long straight magnetic flux tube after the 
dimensional reduction and the proper definition of new angular coordinate 
\cite{dow1,dow2,dow3}.  This is the KK generalization of the Melvin's magnetic 
universe \cite{mel} and is also constructed previously in Ref. \cite{gm}.  
This fact was properly observed in Ref. \cite{sen2}.   The resulting 
solution near the endpoints of the dipole approaches the near-horizon 
solutions for the monopole and the anti-monopole.  In this case, 
it is the net repulsive force due to the external magnetic field that 
keeps the monopole and the anti-monopole pair apart.  
Another way
\footnote{I would like to thank R. Emparan, who made me realize the 
equivalence of the identification under the points under the coordinate 
transformations and the Ehlers-Harrison transformation in the case of 
the KK theory through the email correspondence after the first version of 
this paper appeared in the preprint archive.} 
to introduce a uniform magnetic field into a solution
is through the Ehlers-Harrison transformation \cite{eh}.  The 
Ehlers-Harrison transformation was generalized to the Einstein-Maxwell-dilaton
system (\ref{einmaxdilact}) in Ref. \cite{dow1}.  In Ref. \cite{emp}, this 
generalized Ehlers-Harrison transformation was applied to the dilatonic 
dipole solution (\ref{dipsol}) to construct the solution for the dilatonic 
0-brane/anti-0-brane pair in the external magnetic field.  There, it is 
shown that with the proper strength of external magnetic field the singularity 
of (\ref{dipsol}) at $r=r_+$ disappears and the metric near the end points 
$(r,\theta)=(r_+,\pi)$ and $(r_+,0)$ respectively approaches the near-horizon 
metrics of the dilatonic 0-brane and anti-0-brane.   

In the $\alpha=1$ case, one can generalize the above dipole solution 
in the Einstein-Maxwell-dilaton system to the dyonic dipole solution in the 
Einstein-Maxwell-dilaton-axion system.  The action (\ref{einmaxdilact}) with 
$\alpha=1$ is a special case of the following action for the 
Einstein-Maxwell-dilaton-axion system:
\begin{eqnarray}
S&=&{1\over{2\kappa^2_4}}\int dx^4\sqrt{-g}\left[{\cal R}_g-2(\partial
\phi)^2+{1\over 2}e^{4\phi}(\partial \psi)^2-{1\over 4}e^{-2\alpha\phi}F^2
-{1\over 4}\psi F\tilde{F}\right]
\cr
&=&{1\over{2\kappa^2_4}}\int dx^4\sqrt{-g}\left[{\cal R}_g-{{\partial\lambda
\partial\bar{\lambda}}\over{2(\lambda_2)^2}}-{1\over 4}e^{-2\alpha\phi}F^2
-{1\over 4}\psi F\tilde{F}\right],
\label{dilaxiact}
\end{eqnarray}
where $\psi$ is the axion field, $\lambda=\lambda_1+i\lambda_2\equiv \psi+
ie^{-2\phi}$ is the axion-dilaton field and $\tilde{F}^{\mu\nu}={1\over
{2\sqrt{-g}}}\varepsilon^{\mu\nu\rho\sigma}F_{\rho\sigma}$.  The equations 
of motion for the action (\ref{dilaxiact}) are invariant under the following 
$SL(2,{\bf R})$ transformation \cite{stw}:
\begin{equation}
\lambda\to {{a\lambda+b}\over{c\lambda+d}},\ \ \ \ 
F_{\mu\nu}\to -(c\lambda_1+d)F_{\mu\nu}+c\lambda_2\tilde{F}_{\mu\nu},\ \ \ \ 
g_{\mu\nu}\to g_{\mu\nu},
\label{sl2raxidil}
\end{equation}
where the real numbers $a$, $b$, $c$ and $d$ satisfy $ad-bc=1$.
To construct the dyonic dipole solution to the action (\ref{dilaxiact}), one 
applies the following $SL(2,{\bf R})$ transformation to the electric dipole 
solution (\ref{dipsol}) and (\ref{elecmagdip}) with $\alpha=1$:
\begin{equation}
\lambda\to {{\lambda\cos\delta-\sin\delta}\over{\lambda\sin\delta
+\cos\delta}},\ \ \ \ 
F_{\mu\nu}\to -(\lambda_1\sin\delta+\cos\delta)F_{\mu\nu}
+\lambda_2\sin\delta\tilde{F}_{\mu\nu},\ \ \ \ 
g_{\mu\nu}\to g_{\mu\nu}.
\label{dysl2r}
\end{equation}
The resulting dyonic dipole solution is given by
\begin{eqnarray}
g_{\mu\nu}dx^{\mu}dx^{\nu}&=&-{{r^2-2mr-a^2\cos^2\theta}\over
{r^2-a^2\cos^2\theta}}dt^2
\cr
& &+{{(r^2-2mr-a^2\cos^2\theta)(r^2-a^2\cos^2\theta)}
\over{r^2-2mr+m^2\sin^2\theta-a^2\cos^2\theta}}
\left[{{dr^2}\over{r^2-2mr-a^2}}+d\theta^2\right]
\cr
& &+{{r^2-a^2\cos^2\theta}\over{r^2-2mr-a^2\cos^2\theta}}
(r^2-2mr-a^2)\sin^2\theta d\varphi^2,
\cr
e^{-2\phi}&=&{{(r^2-2mr-a^2\cos^2\theta)(r^2-a^2\cos^2\theta)}
\over{(r^2-2mr-a^2\cos^2\theta)^2\cos^2\delta
+(r^2-a^2\cos^2\theta)^2\sin^2\delta}},
\cr
\psi&=&{{2mr(r^2-mr-a^2\cos^2\theta)\sin 2\delta}\over{(r^2-2mr-
a^2\cos^2\theta)^2\cos^2\delta+(r^2-a^2\cos^2\theta)^2\sin^2\delta}},
\cr
F_{tr}&=&{{4\sqrt{2}mar\cos\delta\cos\theta}\over
{(r^2-a^2\cos^2\theta)^2}},
\cr
F_{t\theta}&=&{{2\sqrt{2}ma\cos\delta\sin\theta(r^2+a^2\cos^2\theta)}
\over{(r^2-a^2\cos^2\theta)^2}},
\cr
F_{r\varphi}&=&{{2\sqrt{2}ma\sin\delta\sin^2\theta(r^2+a^2\cos^2\theta)}
\over{(r^2-2mr-a^2\cos^2\theta)^2}},
\cr 
F_{\theta\varphi}&=&-{{4\sqrt{2}mar\sin\delta\sin\theta\cos\theta
(r^2-2mr-a^2)}\over{(r^2-2mr-a^2\cos^2\theta)^2}}.
\label{dydipol}
\end{eqnarray}
From the expressions for the non-zero components of the field strength 
$F_{\mu\nu}$ in (\ref{dydipol}), one can see that the non-zero components 
of the $U(1)$ field $A_{\mu}$ for the dyonic dipole solution are
\begin{equation}
A_t={{2\sqrt{2}ma\cos\delta\cos\theta}\over{r^2-a^2\cos^2\theta}},\ \ \ \ 
A_{\varphi}=-{{2\sqrt{2}mar\sin\delta\sin^2\theta}\over{r^2-2mr-
a^2\cos^2\theta}}.
\label{dydipolu1}
\end{equation}
This dyonic dipole solution is also constructed in Ref. \cite{muk1}
with the different parametrization of the $SL(2,{\bf R})$ transformation 
matrix.  So, as expected, when the $SO(2)\subset SL(2,{\bf R})$ rotation 
angle is $\delta=\pi/2$, the solution (\ref{dydipol}) becomes magnetic dipole 
solution.  Such ${\bf Z}_2$ transformation corresponds to the 
electric-magnetic duality transformation $A_t\leftrightarrow -A_{\varphi}$,  
$\phi\to -\phi$. The string-frame metric $g^{str}_{\mu\nu}=e^{2\phi}
g_{\mu\nu}$ of the dyonic dipole solution (\ref{dydipol}) is given by
\begin{eqnarray}
g^{str}_{\mu\nu}dx^{\mu}dx^{\nu}&=&[(r^2-2mr-a^2\cos^2\theta)^2\cos^2\delta
+(r^2-a^2\cos^2\theta)^2\sin^2\delta]
\cr
& &\times\left[-{{dt^2}\over{(r^2-a^2\cos^2\theta)^2}}\right.
\cr
& &+{1\over{r^2-2mr+m^2\sin^2\theta-a^2\cos^2\theta}}
\left({{dr^2}\over{r^2-2mr-a^2}}+d\theta^2\right)
\cr
& &\left.+{{r^2-2mr-a^2}\over{(r^2-2mr-a^2\cos^2\theta)^2}}
\sin^2\theta d\varphi^2\right].
\label{strdymet}
\end{eqnarray}
The ADM mass $M$ and the electric $p_{\rm elec}$ and magnetic $p_{\rm mag}$ 
dipole moments of this dyonic dipole solution are
\begin{equation}
M=m=\sqrt{(m\cos\delta)^2+(m\sin\delta)^2},\ \ \ \ 
p_{\rm elec}=ma\cos\delta,\ \ \ \ p_{\rm mag}=-ma\sin\delta.
\label{melmagdip}
\end{equation}

\section{Symmetries of Effective String Theories}

In this section, we discuss the symmetries of the effective string theories 
for the purpose of fixing notations for solutions.  For special values of 
the dilaton coupling $\alpha$, the action (\ref{einmaxdilact}) for the 
Einstein-Maxwell-dilaton system can be embedded as the effective theory of 
string theories.  Thereby, one can study a general dipole solution 
(\ref{dipsol}) with (\ref{elecmagdip}) and its duality related solutions 
within the framework of string theories.  The following effective 
supergravity action for the massless NS-NS sector is common to the 
heterotic, type-IIA and type-IIB string theories:
\begin{equation}
S={1\over{2\kappa^2_{10}}}\int dx^{10}\sqrt{-\hat{G}}\,e^{-\hat{\Phi}}
\left[{\cal R}_{\hat{G}}+\partial_M\hat{\Phi}\partial^M\hat{\Phi}-
{1\over{2\cdot 3!}}\hat{H}_{MNP}\hat{H}^{MNP}\right],
\label{nsnsact}
\end{equation}
where $\hat{\Phi}$ is the $D=10$ dilaton and $\hat{H}_{MNP}$ is the 
field strength of the NS-NS 2-form potential $\hat{B}_{MN}$.  

If one assumes the KK Ansatz for the metric of the form 
$(\hat{G}_{MN})={\rm diag}(G_{\bar{\mu}\bar{\nu}},\delta_{\bar{m}\bar{n}})$ 
($\bar{\mu},\bar{\nu}=0,1,...,5$; $\bar{m},\bar{n}=6,...,9$) and let 
$\hat{B}_{\bar{\mu}\bar{n}}=0=\hat{B}_{\bar{m}\bar{n}}$, then the effective 
action (\ref{nsnsact}) is compactified to the following $D=6$ action:
\begin{equation}
S={1\over{2\kappa^2_6}}\int dx^6\sqrt{-G}\,e^{-\Phi}\left[{\cal R}_G+
\partial_{\bar{\mu}}\Phi\partial^{\bar{\mu}}\Phi-
{1\over{2\cdot 3!}}H_{\bar{\mu}\bar{\nu}\bar{\rho}}
H^{\bar{\mu}\bar{\nu}\bar{\rho}}\right],
\label{6dact}
\end{equation}
where $\Phi=\hat{\Phi}$ is the $D=6$ dilaton and $H_{\bar{\mu}\bar{\nu}
\bar{\rho}}$ is the field strength of the NS-NS 2-form potential 
$B_{\bar{\mu}\bar{\nu}}=\hat{B}_{\bar{\mu}\bar{\nu}}$.  Note, such 
dimensional reduction is a consistent truncation of the $D=10$ 
superstring effective action.  

We further compactify the $D=6$ action (\ref{6dact}) down to 
$D=4$ by using the following Ans\"atze for the fields:
\begin{eqnarray}
G_{\bar{\mu}\bar{\nu}}&=&\left(\matrix{e^{\eta}g_{\mu\nu}+A^m_{\mu}
A^n_{\nu}G_{mn} & A^m_{\mu}G_{mn}\cr A^n_{\nu}G_{mn} & G_{mn}}\right),
\cr
B_{\bar{\mu}\bar{\nu}}&=&\left(\matrix{B_{\mu\nu}+{1\over 2}(A^m_{\mu}
B_{m\nu}-B_{\mu n}A^n_{\nu}) & B_{\mu n}+A^m_{\mu}B_{mn}\cr 
B_{m\nu}+B_{mn}A^n_{\nu} & B_{mn}}\right),
\label{6d4dansat}
\end{eqnarray}
where the indices run as $\mu,\nu=0,...,3$ and $m,n=4,5$, and $e^{\eta}
\equiv e^{\Phi}/\sqrt{\det G_{mn}}=e^{\Phi+\sigma}$ is the $D=4$ dilaton.  
We parametrize the scalars $G_{mn}$ and $B_{mn}$ in the following way:
\begin{equation}
G_{mn}=e^{\rho-\sigma}\left(\matrix{e^{-2\rho}+c^2 & -c\cr -c & 
1}\right),\ \ \ \ \ 
B_{mn}=b\varepsilon_{mn},
\label{scalmet}
\end{equation}
and we Hodge-dualize the field strength $H_{\mu\nu\rho}$ to define the 
$D=4$ axion $a$:
\begin{equation}
\varepsilon^{\mu\nu\rho\sigma}\partial_{\sigma}a=\sqrt{-g}e^{-\eta}
g^{\mu\sigma}g^{\nu\lambda}g^{\rho\tau}H_{\sigma\lambda\tau}.
\label{axion}
\end{equation}
In terms of the real scalars defined above, we further define the 
following complex scalars:
\begin{equation}
S=S_1+iS_2\equiv a+ie^{-\eta},\ \ \ 
T=T_1+iT_2\equiv b+ie^{-\sigma},\ \ \ 
U=U_1+iU_2\equiv c+ie^{-\rho},
\label{cmplxsc}
\end{equation}
where $S$ is the dilaton-axion field, and $T$ and $U$ are respectively 
the K\"ahler structure and the complex structure of $T^2$ and can be 
expressed as the following $SL(2,{\bf R})$ matrices:
\begin{equation}
{\cal M}_T\equiv {1\over{T_2}}\left(\matrix{1 & T_1\cr T_1 & 
|T|^2}\right), \ \ \ \ \ 
{\cal M}_U\equiv {1\over{U_2}}\left(\matrix{1 & U_1\cr U_1 & 
|U|^2}\right).
\label{tumats}
\end{equation}

The final form of the effective action in $D=4$ is then \cite{dlr}
\begin{eqnarray}
S&=&{1\over{2\kappa^2_4}}\int dx^4\sqrt{-g}\left[{\cal R}_g
+{1\over 4}{\rm Tr}(\partial_{\mu}{\cal M}^{-1}_T\partial^{\mu}
{\cal M}_T)+{1\over 4}{\rm Tr}(\partial_{\mu}{\cal M}^{-1}_U
\partial^{\mu}{\cal M}_U)\right.
\cr
& &\left.-{1\over{2(S_2)^2}}\partial_{\mu}S\partial^{\mu}\bar{S}
-{1\over 4}S_2{\cal F}^{\rm T}_{\mu\nu}({\cal M}_T\otimes{\cal M}_U)
{\cal F}^{\mu\nu}\right],
\label{4dact}
\end{eqnarray}
where ${\cal F}_{\mu\nu}=({\cal F}^i_{\mu\nu})$ are the field strengths 
of the $U(1)$ gauge fields ${\cal A}^i_{\mu}$ ($i=1,...,4$) defined as 
${\cal A}^1_{\mu}=B_{4\mu}$, ${\cal A}^2_{\mu}=B_{5\mu}$, ${\cal 
A}^3_{\mu}=A^5_{\mu}$ and ${\cal A}^4_{\mu}=A^4_{\mu}$.  The action 
(\ref{4dact}) is manifestly invariant under the following 
$SL(2,{\bf R})_T\times SL(2,{\bf R})_U$ $T$-duality transformation \cite{dlr}:
\begin{equation}
{\cal M}_T\to \omega^{\rm T}_T{\cal M}_T\omega_T,\ \ \ 
{\cal M}_U\to \omega^{\rm T}_U{\cal M}_U\omega_U,\ \ \ 
{\cal F}_{\mu\nu}\to (\omega^{-1}_T\otimes\omega^{-1}_U){\cal 
F}_{\mu\nu},
\label{sl2rsl2r}
\end{equation}
where $\omega_{T,U}\in SL(2,{\bf R})_{T,U}$.  In addition, the theory has 
an on-shell symmetry under the following $SL(2,{\bf R})_S$ $S$-duality 
transformation \cite{dlr}:
\begin{equation}
S\to{{aS+b}\over{cS+d}},\ \ \ \ 
\left(\matrix{{\cal F}^i_{\mu\nu}\cr \tilde{\cal F}^i_{\mu\nu}}\right)
\to \omega^{-1}_S\left(\matrix{{\cal F}^i_{\mu\nu}\cr 
\tilde{\cal F}^i_{\mu\nu}}\right);\ \ \ \ \ \omega_S=\left(\matrix{a&b\cr 
c&d}\right)\in SL(2,{\bf R})_S,
\label{sdualtr}
\end{equation}
where $\tilde{\cal F}^i_{\mu\nu}$ is the dual to the $U(1)$ field strength 
${\cal F}^i_{\mu\nu}$.  

In particular, when the real parts of all the complex scalars (\ref{cmplxsc}) 
are zero, the action (\ref{4dact}) takes the following form:
\begin{eqnarray}
S&=&{1\over{2\kappa^2_4}}\int dx^4\sqrt{-g}\left[{\cal R}_g
-{1\over 2}\left\{(\partial\eta)^2+(\partial\sigma)^2+(\partial\rho)^2
\right\}\right.
\cr
& &\left.-{{e^{-\eta}}\over 4}\left\{e^{\sigma+\rho}({\cal F}^1)^2
+e^{\sigma-\rho}({\cal F}^2)^2+e^{-\sigma+\rho}({\cal F}^3)^2+
e^{-\sigma-\rho}({\cal F}^4)^2\right\}\right].
\label{scalact}
\end{eqnarray}
When only $n$ of the gauge fields ${\cal A}^i_{\mu}$ (or 
$\tilde{\cal A}^i_{\mu}$) are non-zero and equal, the action (\ref{scalact}) 
can be transformed to the form of the Einstein-Maxwell-dilaton action 
(\ref{einmaxdilact}) with the dilaton coupling $\alpha=\sqrt{(4-n)/n}$, 
after the field redefinition.  Thereby, the dipole solution (\ref{dipsol}) 
is embedded as a solution of the effective string theory when the dilaton 
coupling takes the value $\alpha=\sqrt{(4-n)/n}=\sqrt{3},1,1/\sqrt{3},0$.  
The dipole solution with $\alpha=\sqrt{(4-n)/n}$ is therefore interpreted as 
the bound state of $n$ ``fundamental'' dipoles with the dilaton coupling 
$\alpha=\sqrt{3}$ (cf. \cite{rahm,rduff}).  

When the action (\ref{nsnsact}) is regarded as the bosonic effective action 
for the NS-NS sector of the type-IIB string theory, one can apply the 
$SL(2,{\bf R})$ $S$-duality transformation of the type-IIB theory to the 
dipole solutions carrying the dipole moments of the $U(1)$ fields originated 
from the NS-NS 2-form field to obtain the dipole solutions carrying the 
dipole moments of the $U(1)$ fields originated from the RR 2-form field.  
The bosonic part of the effective action for the type-IIB theory is given by 
\cite{bbo}
\begin{eqnarray}
S_{IIB}&=&{1\over{2\kappa^2_{10}}}\int dx^{10}\sqrt{-\hat{G}}\left[
e^{-\hat{\Phi}}\left\{{\cal R}_{\hat{G}}+(\partial\hat{\Phi})^2
-{1\over{2\cdot 3!}}(\hat{H}^{(1)})^2\right\}-{1\over 2}(\partial
\hat{\chi})^2\right.
\cr
&-&\left.{1\over{2\cdot 3!}}(\hat{H}^{(2)}
-\hat{\chi}\hat{H}^{(1)})^2-{5\over 6}(\hat{F})^2-{1\over{2^5\cdot 3^3
\sqrt{-\hat{G}}}}\varepsilon^{ij}\varepsilon\hat{D}\hat{H}^{(i)}
\hat{H}^{(i)}\right],
\label{iibact}
\end{eqnarray}
where $\hat{H}^{(1)}$ [$\hat{H}^{(2)}$] is the field strength of the 
NS-NS [RR] 2-form potential $\hat{B}^{(1)}$ [$\hat{B}^{(2)}$], 
$\hat{\chi}$ is the RR 0-form field and $\hat{F}$ is the field strength 
of the RR 4-form potential $\hat{D}$.  The $SL(2,{\bf R})$ symmetry of 
the effective action is manifest in the Einstein-frame (with the spacetime 
metric $\hat{G}^E_{MN}$).  After applying the Weyl-scaling transformation 
$\hat{G}^E_{MN}=e^{-{\hat{\Phi}\over 4}}\hat{G}_{MN}$, one obtains the 
following Einstein-frame action:
\begin{eqnarray}
S_{IIB}&=&{1\over{2\kappa^2_{10}}}\int dx^{10}\sqrt{-\hat{G}^E}\left[
{\cal R}_{\hat{G}^E}+{1\over 4}{\rm Tr}(\partial_M\hat{\cal M}
\partial^M\hat{\cal M}^{-1})-{1\over{2\cdot 3!}}\hat{H}^{(i)}\hat{\cal M}_{ij}
\hat{H}^{(j)}\right]
\cr
& &\left.-{5\over 6}(\hat{F})^2-{1\over{2^5\cdot 3^3
\sqrt{-\hat{G}^E}}}\varepsilon^{ij}\varepsilon\hat{D}\hat{H}^{(i)}
\hat{H}^{(j)}\right],
\label{einiibact}
\end{eqnarray}
where a $2\times 2$ real matrix $\hat{\cal M}$ is defined in terms of the 
complex scalar $\hat{\lambda}=\hat{\chi}+ie^{-{\hat{\Phi}\over 2}}$ as
\begin{equation}
\hat{\cal M}={1\over{{\cal I}{\rm m}\,\hat{\lambda}}}\left(
\matrix{|\hat{\lambda}|^2 & {\cal R}{\rm e}\,\hat{\lambda}\cr
{\cal R}{\rm e}\,\hat{\lambda} & 1}\right).
\label{sl2rscal}
\end{equation}
The action (\ref{einiibact}) is manifestly invariant under the following 
$SL(2,{\bf R})$ transformation \cite{hw,sch}:
\begin{equation}
\left(\matrix{\hat{H}^{(1)}\cr \hat{H}^{(2)}}\right)\to (\omega^{\rm T})^{-1}
\left(\matrix{\hat{H}^{(1)}\cr \hat{H}^{(2)}}\right),\ \ \ 
\hat{\cal M}\to \omega\hat{\cal M}\omega^{\rm T},\ \ \ \ 
\omega\in SL(2,{\bf R}).
\label{sl2rtran}
\end{equation}

To compactify the string-frame action (\ref{iibact}) down to $D=6$, similarly 
to the previous case, we use the simplified field Ansatz where (spacetime and 
internal space) mixing components of fields are zero and only internal space 
components of the RR 4-form potential $\hat{D}$ are non-zero.  However, 
unlike the previous case, we take the KK Ansatz for the $D=10$ metric 
$\hat{G}_{MN}$ to be $(\hat{G}_{MN})=(G_{\bar{\mu}\bar{\nu}},e^{\bar{G}}
\delta_{\bar{m}\bar{n}})$.  The resulting $D=6$ action is as follows 
\cite{bbo}:
\begin{eqnarray}
S&=&{1\over{2\kappa^2_6}}\int dx^6\sqrt{-G}\left[e^{-\Phi}\left\{{\cal R}_G
+(\partial\Phi)^2-(\partial\bar{G})^2-{1\over{2\cdot 3!}}(H^{(1)})^2
\right\}-{1\over 2}e^{2\bar{G}}(\partial\chi)^2\right.
\cr
& &\left.-{1\over{2\cdot 3!}}e^{2\bar{G}}(\hat{H}^{(2)}-\chi\hat{H}^{(1)})^2
-{1\over{72}}e^{-2\bar{G}}(\partial D)^2+{1\over{72}}H^{(i)}{\cal L}_{ij}\,
\star H^{(j)}\right],
\label{iib6dact}
\end{eqnarray}
where $\Phi=\hat{\Phi}$ is the $D=6$ dilaton, $\chi=\hat{\chi}$ is the 
$D=6$ RR 0-form field, $D\equiv\varepsilon^{\bar{m}\bar{n}\bar{p}\bar{q}}
\hat{D}_{\bar{m}\bar{n}\bar{p}\bar{q}}$, $(\star H^{(i)})_{\bar{\mu}
\bar{\nu}\bar{\rho}}={1\over{3!\sqrt{-G}}}\varepsilon_{\bar{\mu}\bar{\nu}
\bar{\rho}\bar{\alpha}\bar{\beta}\bar{\gamma}}H^{(i)\,\bar{\alpha}\bar{\beta}
\bar{\gamma}}$ and ${\cal L}=\left(\matrix{0&1\cr -1&0}\right)$.  The symmetry 
of the $D=6$ theory is manifest in the following Einstein-frame action, which 
is achieved by the Weyl-scaling $G^E_{\bar{\mu}\bar{\nu}}=e^{-{\Phi\over 2}}
G_{\bar{\mu}\bar{\nu}}$:
\begin{eqnarray}
S&=&{1\over{2\kappa^2_6}}\int dx^6\sqrt{-G^E}\left[{\cal R}_{G^E}
+{{\partial\lambda\partial\bar{\lambda}}\over{2(\lambda_2)^2}}
+{{\partial\tau\partial\bar{\tau}}\over{2(\tau_2)^2}}\right.
\cr
& &\left.\ \ \ -{{\tau_2}\over 9}H^{(i)}\hat{\cal M}_{ij}H^{(j)}
+{{\tau_1}\over 9}H^{(i)}{\cal L}_{ij}\,\star H^{(j)}\right],
\label{6diibeinact}
\end{eqnarray}
where the complex scalars $\lambda$ and $\tau$ are defined as
\begin{eqnarray}
\lambda&=&\lambda_1+i\lambda_2=\chi+ie^{-{{\Phi}\over 2}},
\cr
\tau&=&\tau_1+i\tau_2={1\over 8}D+i{3\over 4}e^{2G},
\label{cmpxsc}
\end{eqnarray}
where $G\equiv\bar{G}-\Phi/4$.  In addition to the $SL(2,{\bf R})$ $S$-duality 
symmetry (\ref{sl2rtran}) of the $D=10$ Einstein-frame action 
(\ref{einiibact}), the action (\ref{6diibeinact}) also has the on-shell 
symmetry under the following $SL(2,{\bf R})_{EM}$ electric-magnetic 
transformation 
\cite{bbo}:
\begin{equation}
\tau\to{{p\tau+q}\over{r\tau+s}},\ \ \ 
H^{(i)}\to (r\tau_1+s)H^{(i)}+r\tau_2({\cal L}\hat{\cal M})_{ij}\,\star 
H^{(j)},\ \ \ \ \ ps-qr=1.
\label{2l2remtrn}
\end{equation}
Using this electric-magnetic transformation, one can construct other $D=4$ 
dyonic dipole solutions from the dipole solutions presented in this paper.  

\section{Fundamental Dipole Solutions}

In this section, we write down ``fundamental'' dipole solutions, which we 
define as dipole solutions with either electric or magnetic component of 
only one $U(1)$ gauge field non-zero.  In this case, the corresponding 
$D=4$ string effective action (\ref{4dact}) can be transformed to the form 
(\ref{einmaxdilact}) with the dilaton coupling $\alpha=\sqrt{3}$ through 
field redefinition.  So, the Einstein-frame metric $g_{\mu\nu}$ is given by 
(\ref{dipsol}) with $\alpha=\sqrt{3}$, which can also be check by applying 
the various ${\bf Z}_2$ subset duality transformations to say KK dipole 
solutions:
\begin{eqnarray}
g_{\mu\nu}dx^{\mu}dx^{\nu}&=&-\sqrt{{r^2-2mr-a^2\cos^2\theta}\over
{r^2-a^2\cos^2\theta}}dt^2
\cr
& &+\sqrt{(r^2-2mr-a^2\cos^2\theta)(r^2-a^2\cos^2\theta)}
\left[{{dr^2}\over{r^2-2mr-a^2}}+d\theta^2\right]
\cr
& &+\sqrt{{r^2-a^2\cos^2\theta}\over{r^2-2mr-a^2\cos^2\theta}}
(r^2-2mr-a^2)\sin^2\theta d\varphi^2.
\label{fundimet}
\end{eqnarray}
In the following, we show the explicit expressions for the other fields for 
various cases of fundamental dipoles.

The case where only the magnetic component of the KK $U(1)$ gauge 
field ${\cal A}^3_{\mu}=A^5_{\mu}$ is non-zero corresponds to the $D=5$ 
KK magnetic dipole solution constructed in Ref. \cite{gper}.  In terms of 
the effective string theory field parametrization, the solution is 
rewritten as follows:
\begin{eqnarray}
\eta&=&\sigma=-\rho=-{1\over 2}\ln{{r^2-2mr-a^2\cos^2\theta}\over
{r^2-a^2\cos^2\theta}},
\cr
{\cal A}^3_{\varphi}&=&A^5_{\varphi}={{2mar\sin^2\theta}\over
{r^2-2mr-a^2\cos^2\theta}}.
\label{kkdipol}
\end{eqnarray}
When ${\cal A}^3_{\mu}=A^5_{\mu}$ is electric, the corresponding solution 
(\ref{dipsol}) and (\ref{elecmagdip}) can be rewritten in terms of the fields 
of the effective string theory as
\begin{eqnarray}
\eta&=&\sigma=-\rho={1\over 2}\ln{{r^2-2mr-a^2\cos^2\theta}\over
{r^2-a^2\cos^2\theta}},
\cr
{\cal A}^3_t&=&A^5_t={{2ma\cos\theta}\over{r^2-a^2\cos^2\theta}},
\label{kkelecdipol}
\end{eqnarray}
with the real scalars having the opposite signs to the magnetic case.  
This is the KK electric dipole solution, which is the electric-magnetic dual 
to the solution in Ref. \cite{gper}.  The remaining KK dipole solutions whose 
KK $U(1)$ fields come from the other circle are obtained by applying the 
${\bf Z}_2$ subset of the $SL(2,{\bf R})_U$ transformation (\ref{sl2rsl2r}) 
to the above solutions (\ref{kkdipol}) and (\ref{kkelecdipol}).  Note that 
the ${\bf Z}_2$ subset of $SL(2,{\bf R})_U$ maps the KK [winding] $U(1)$ 
field of one circle to the KK [winding] $U(1)$ field of the other circle, 
while changing the sign of the real scalar $\rho$ of the complex structure 
$U$.  So, the remaining KK dipole solutions are
\begin{eqnarray}
\eta&=&\sigma=\rho=-{1\over 2}\ln{{r^2-2mr-a^2\cos^2\theta}\over
{r^2-a^2\cos^2\theta}},
\cr
{\cal A}^4_{\varphi}&=&A^4_{\varphi}={{2mar\sin^2\theta}\over
{r^2-2mr-a^2\cos^2\theta}},
\label{kkdipol2}
\end{eqnarray}
and 
\begin{eqnarray}
\eta&=&\sigma=\rho={1\over 2}\ln{{r^2-2mr-a^2\cos^2\theta}\over
{r^2-a^2\cos^2\theta}},
\cr
{\cal A}^4_t&=&A^4_t={{2ma\cos\theta}\over{r^2-a^2\cos^2\theta}}.
\label{kkelecdipol2}
\end{eqnarray}

The ${\bf Z}_2$ subset of the $SL(2,{\bf R})_T$ transformation 
(\ref{sl2rsl2r}) maps the KK electric [magnetic] $U(1)$ field of 
one circle to the winding electric [magnetic] $U(1)$ field of the other 
circle and vice versa, while changing the sign of the real scalar $\sigma$ 
of the K\"ahler structure $T$.  By applying this ${\bf Z}_2$ transformation 
to the above KK dipole solutions, one obtains the winding dipole 
(or $H$ dipole) solutions.  The magnetic $H$ dipole solutions are
\begin{eqnarray}
\eta&=&-\sigma=-\rho=-{1\over 2}\ln{{r^2-2mr-a^2\cos^2\theta}\over
{r^2-a^2\cos^2\theta}},
\cr
{\cal A}^1_{\varphi}&=&B_{4\,\varphi}={{2mar\sin^2\theta}\over
{r^2-2mr-a^2\cos^2\theta}},
\label{magwindip1}
\end{eqnarray}
and
\begin{eqnarray}
\eta&=&-\sigma=\rho=-{1\over 2}\ln{{r^2-2mr-a^2\cos^2\theta}\over
{r^2-a^2\cos^2\theta}},
\cr
{\cal A}^2_{\varphi}&=&B_{5\,\varphi}={{2mar\sin^2\theta}\over
{r^2-2mr-a^2\cos^2\theta}}.
\label{magwindip2}
\end{eqnarray}
The electric $H$ dipole solutions are
\begin{eqnarray}
\eta&=&-\sigma=-\rho={1\over 2}\ln{{r^2-2mr-a^2\cos^2\theta}\over
{r^2-a^2\cos^2\theta}},
\cr
{\cal A}^1_t&=&B_{4\,t}={{2ma\cos\theta}\over{r^2-a^2\cos^2\theta}},
\label{elecwindip1}
\end{eqnarray}
and
\begin{eqnarray}
\eta&=&-\sigma=\rho={1\over 2}\ln{{r^2-2mr-a^2\cos^2\theta}\over
{r^2-a^2\cos^2\theta}},
\cr
{\cal A}^2_t&=&B_{5\,t}={{2ma\cos\theta}\over{r^2-a^2\cos^2\theta}}.
\label{elecwindip2}
\end{eqnarray}
These $H$ dipole solutions can also be obtained by applying the 
${\bf Z}_2$ subset of the $S$-duality transformation (\ref{sdualtr}), 
which maps the KK electric [magnetic] field of one circle 
to the $H$ magnetic [electric] field of the same circle and vice 
versa and changes the sign of the real scalar $\eta$ of the dilaton-axion 
field $S$, to the KK dipole solutions.  

The parametrization of scalars of the above fundamental NS-NS dipole 
solutions in terms of the internal metric $G_{mn}$ is achieved by 
using Eq. (\ref{6d4dansat}), namely $(G_{mn})={\rm diag}(e^{-\rho-\sigma},
e^{\rho-\sigma})$.  And of course the $D=4$ dilaton is $\eta$. 

From the expressions for the $D=4$ dilaton $\eta$ in the above NS-NS  
fundamental dipole solutions, one can see that the string-frame metric 
$g^{str}_{\mu\nu}=e^{\eta}g_{\mu\nu}$ for all the electric NS-NS 
fundamental dipole solutions has the following form:
\begin{eqnarray}
g^{\rm str}_{\mu\nu}dx^{\mu}dx^{\nu}&=&-\left({r^2-2mr-a^2\cos^2\theta}
\over{r^2-a^2\cos^2\theta}\right)dt^2
\cr
& &+(r^2-2mr-a^2\cos^2\theta)\left[{{dr^2}\over{r^2-2mr-a^2}}+d\theta^2\right]
\cr
& &+(r^2-2mr-a^2)\sin^2\theta d\varphi^2,
\label{strnselec}
\end{eqnarray}
and for all the magnetic NS-NS fundamental dipole solutions the 
string-frame metric is given by:
\begin{eqnarray}
g^{\rm str}_{\mu\nu}dx^{\mu}dx^{\nu}&=&-dt^2+(r^2-a^2\cos^2\theta)
\left[{{dr^2}\over{r^2-2mr-a^2}}+d\theta^2\right]
\cr
& &+{{r^2-a^2\cos^2\theta}\over{r^2-2mr-a^2\cos^2\theta}}
(r^2-2mr-a^2)\sin^2\theta d\varphi^2.
\label{strnsmag}
\end{eqnarray}

The ${\bf Z}_2$ subset of the $SL(2,{\bf R})$ $S$-duality transformation 
(\ref{sl2rtran}) of the type-IIB theory interchanges the NS-NS 2-form 
potential $B^{(1)}_{MN}=\hat{B}_{MN}$ and the RR 2-form potential 
$B^{(2)}_{MN}$.  So, by applying this ${\bf Z}_2$ transformation to the $H$ 
dipole solutions, one obtains the dipole solutions charged with respect to the 
$U(1)$ field ${\cal B}^a_{\mu}\equiv \hat{B}^{(2)}_{3+a,\mu}$ originated 
from the RR 2-form potential $\hat{B}^{(2)}_{MN}$, which we name as 
D dipoles. Note that in the $H$ dipole solutions in Eqs. (\ref{magwindip1}) 
$-$ (\ref{elecwindip2}) the scalar $\bar{G}$ is zero, i.e. 
$\hat{G}_{\bar{m}\bar{n}}=\delta_{\bar{m}\bar{n}}$.  Under the $SL(2,{\bf R})$ 
transformation (\ref{sl2rtran}) to these $H$ dipole solutions, the scalar 
$\bar{G}$ is induced in such a way that the combination $G=\bar{G}-\Phi/4$ is 
invariant.  Such ${\bf Z}_2$ transformed solutions have the same 
Einstein-frame metric as the NS-NS ``fundamental'' dipole solutions in the 
above, but the $D=4$ dilaton $\phi\equiv\hat{\Phi}-{1\over 2}\ln\,{\rm det}
\,g_{ij}$ and the internal metric $g_{ij}\equiv\hat{G}_{3+i,3+j}$ ($i,j=
1,...,6$) differ.  The solutions are 
\begin{eqnarray}
g_{11}&=&\sqrt{{r^2-2mr-a^2\cos^2\theta}\over{r^2-a^2\cos^2\theta}},\ \ \ \ 
g_{ii}=\sqrt{{r^2-a^2\cos^2\theta}\over{r^2-2mr-a^2\cos^2\theta}}\ \ \ 
(i\neq 1),
\cr
{\cal B}^1_t&=&{{2ma\cos\theta}\over{r^2-a^2\cos^2\theta}},
\label{elecrrdip}
\end{eqnarray}
for the electric case, and 
\begin{eqnarray}
g_{11}&=&\sqrt{{r^2-a^2\cos^2\theta}\over{r^2-2mr-a^2\cos^2\theta}},\ \ \ \ 
g_{ii}=\sqrt{{r^2-2mr-a^2\cos^2\theta}\over{r^2-a^2\cos^2\theta}}\ \ \ 
(i\neq 1),
\cr
{\cal B}^1_{\varphi}&=&{{2mar\sin^2\theta}\over{r^2-2mr-a^2\cos^2\theta}}, 
\label{magrrdip}
\end{eqnarray}
for the magnetic case.  The $D=4$ dilaton is $\phi=0$ for both cases and 
therefore the Einstein- and string-frame metrics have the same form.

\subsection{Higher-dimensional Embeddings}

By uplifting the fundamental dipole solutions constructed in the above 
to $D=10$, one can obtain the supergravity solutions for brane/anti-brane 
pairs in $D=10$.  Although for such solutions some of the transverse 
directions are delocalized, one may be able to learn about the complete 
solutions from such delocalized solutions.  Of course, one can compactify 
such delocalized transverse directions of the solutions for the 
brane/anti-brane pairs and their bound states presented in this subsection 
and the following section to obtain completely localized solutions in $D<10$.  

\subsubsection{F-string/anti-F-string pair}

By uplifting the electric $H$ dipole solution (\ref{elecwindip1}) 
or (\ref{elecwindip2}) to $D=10$, one obtains the following supergravity 
solution for the fundamental string (F-string) and anti-fundamental 
string pair:
\begin{eqnarray}
ds^2_{10}&=&H^{-1}\left[-dt^2+(dx^1)^2\right]
+(dx^2)^2+\cdots+(dx^6)^2
\cr
& &+(r^2-2mr-a^2\cos^2\theta)\left[{{dr^2}\over{r^2-2mr-a^2}}
+d\theta^2\right]
\cr
& &+(r^2-2mr-a^2)\sin^2\theta d\varphi^2,
\cr
\hat{B}_{x^1t}&=&{{2ma\cos\theta}\over{r^2-a^2\cos^2\theta}},
\ \ \ \ \ \ 
e^{\hat{\Phi}}=H^{-1},
\label{10df}
\end{eqnarray}
where the ``modified'' harmonic function is
\begin{equation}
H={{r^2-a^2\cos^2\theta}\over{r^2-2mr-a^2\cos^2\theta}}.
\label{fharm}
\end{equation}

\subsubsection{D2-brane/anti-D2-brane pair}

The following supergravity solution for the D2-brane and the anti-D2-brane 
pair can be constructed by uplifting the F-string/anti-F-string pair 
solution (\ref{10df}) to $D=11$ and then compactifying one of the transverse 
directions of the resulting M2-brane/anti-M2-brane pair solution on $S^1$:
\begin{eqnarray}
ds^2_{10}&=&H^{-{1\over 2}}\left[-dt^2+(dx^1)^2
+(dx^2)^2\right]+H^{1\over 2}\left[(dx^3)^2+\cdots+(dx^6)^2
\right.
\cr
& &\left.+(r^2-2mr-a^2\cos^2\theta)
\left({{dr^2}\over{r^2-2mr-a^2}}+d\theta^2\right)\right.
\cr
& &\left.+(r^2-2mr-a^2)\sin^2\theta d\varphi^2\right],
\cr
\hat{B}_{tx^1x^2}&=&-{{2ma\cos\theta}\over{r^2-a^2\cos^2\theta}},
\ \ \ \ \ \  
e^{\hat{\Phi}}=H^{1\over 2},
\label{10dd2}
\end{eqnarray}
where $\hat{B}_{MNP}$ is the 3-form potential in the RR-sector of 
the type-IIA string theory.

\subsubsection{D-string/anti-D-string pair}

By uplifting the electric dipole solution (\ref{elecrrdip}) charged under 
the RR $U(1)$ field ${\cal B}^1_{\mu}$, one obtains the following 
supergravity solution for D-string and anti-D-string pair:
\begin{eqnarray}
ds^2_{10}&=&H^{-{1\over 2}}[-dt^2+(dx^1)^2]
+H^{1\over 2}[(dx^2)^2+\cdots+(dx^6)^2
\cr
& &+(r^2-2mr-a^2\cos^2\theta)\left({{dr^2}\over{r^2-2mr-a^2}}
+d\theta^2\right)
\cr
& &\left.+(r^2-2mr-a^2)\sin^2\theta d\varphi^2\right],
\cr
\hat{B}^{(2)}_{x^1t}&=&{{2ma\cos\theta}\over{r^2-a^2\cos^2\theta}},
\ \ \ \ \ \ 
e^{\hat{\Phi}}=H.
\label{10dd}
\end{eqnarray}

\subsubsection{NS5-brane/anti-NS5-brain pair}

By uplifting the magnetic $H$ dipole solution (\ref{magwindip1}) or 
(\ref{magwindip2}) to $D=10$, one obtains the following supergravity 
solution for the NS5-brane and the anti-NS5-brane pair:
\begin{eqnarray}
ds^2_{10}&=&-dt^2+(dx^1)^2+\cdots+(dx^5)^2+H[(dx^6)^2
\cr
& &+(r^2-2mr-a^2\cos^2\theta)\left({{dr^2}\over{r^2-2mr-a^2}}
+d\theta^2\right)
\cr
& &\left.+(r^2-2mr-a^2)\sin^2\theta d\varphi^2\right],
\cr
\hat{B}_{x^6\varphi}&=&{{2mar\sin^2\theta}\over{r^2-2mr-a^2\cos^2\theta}},
\ \ \ \ \ 
e^{\hat{\Phi}}=H.
\label{10dns5}
\end{eqnarray}

\subsubsection{D4-brain/anti-D4-brane pair}

The following supergravity solution for the D4-brane and the 
anti-D4-brane pair can be obtained by first uplifting the 
NS5-brane/anti-NS5-brain pair solution (\ref{10dns5}) to $D=11$ 
and then compactifying one of the longitudinal directions of the 
resulting M5-brane/anti-M5-brane solution on $S^1$:
\begin{eqnarray}
ds^2_{10}&=&H^{-{1\over 2}}\left[-dt^2+(dx^1)^2+\cdots
+(dx^4)^2\right]+H^{1\over 2}\left[(dx^5)^2+(dx^6)^2\right.
\cr
& &+(r^2-2mr-a^2\cos^2\theta)\left({{dr^2}\over{r^2-2mr-a^2}}
+d\theta^2\right)
\cr
& &\left.+(r^2-2mr-a^2)\sin^2\theta d\varphi^2\right],
\cr
\hat{B}_{\varphi x^5x^6}&=&-{{2mar\sin^2\theta}\over
{r^2-2mr-a^2\cos^2\theta}},\ \ \ \ \ 
e^{\hat{\Phi}}=H^{-{1\over 2}}.
\label{10dd4}
\end{eqnarray}

\subsubsection{D5-brain/anti-D5-brane pair}

By uplifting the magnetic D dipole solution (\ref{magrrdip}) to $D=10$, 
one obtains the following supergravity solution for the D5-brane and the 
anti-D5-brane pair:
\begin{eqnarray}
ds^2_{10}&=&H^{-{1\over 2}}\left[-dt^2+(dx^1)^2+\cdots
+(dx^5)^2\right]+H^{1\over 2}\left[(dx^6)^2\right.
\cr
& &+(r^2-2mr-a^2\cos^2\theta)\left({{dr^2}\over{r^2-2mr-a^2}}
+d\theta^2\right)
\cr
& &\left.+(r^2-2mr-a^2)\sin^2\theta d\varphi^2\right],
\cr
\hat{B}^{(2)}_{x^6\varphi}&=&{{2mar\sin^2\theta}\over
{r^2-2mr-a^2\cos^2\theta}},\ \ \ \ \ \ 
e^{\hat{\Phi}}=H^{-1}.
\label{10dd5}
\end{eqnarray}

\subsubsection{pp-wave/anti-pp-wave pair}

By uplifting the electric KK dipole solution (\ref{kkelecdipol}) or 
(\ref{kkelecdipol2}) to $D=10$, one obtains the following supergravity 
solution for the pp-wave and anti-pp-wave pair:
\begin{eqnarray}
ds^2_{10}&=&-H^{-1}dt^2+H(dx^1+{{2ma\cos\theta}\over
{r^2-a^2\cos^2\theta}}dt)^2
\cr
& &+(dx^2)^2+\cdots+(dx^6)^2
\cr
& &+(r^2-2mr-a^2\cos^2\theta)
\left[{{dr^2}\over{r^2-2mr-a^2}}+d\theta^2\right]
\cr
& &+(r^2-2mr-a^2)\sin^2\theta d\varphi^2.
\label{10dpp}
\end{eqnarray}

\subsubsection{D0-brane/anti-D0-brane pair}

To obtain the following supergravity solution for the D0-brane and the 
anti-D0-brane pair, one first embed the pp-wave/anti-pp-wave solution 
(\ref{10dpp}) into $D=11$ supergravity and then compactify along the 
longitudinal direction of the $D=11$ pp-wave:
\begin{eqnarray}
ds^2_{10}&=&-H^{-{1\over 2}}dt^2+H^{1\over 2}
\left[(dx^1)^2+\cdots+(dx^6)^2\right.
\cr
& &+(r^2-2mr-a^2\cos^2\theta)\left({{dr^2}\over{r^2-2mr-a^2}}
+d\theta^2\right)
\cr
& &\left.+(r^2-2mr-a^2)\sin^2\theta d\varphi^2\right],
\cr
\hat{A}_t&=&{{2ma\cos\theta}\over{r^2-a^2\cos^2\theta}},\ \ \ \ \ 
e^{\hat{\Phi}}=H^{3\over 2},
\label{10dd0}
\end{eqnarray}
where $\hat{A}_M$ is the 1-form field in the RR-sector of the type-IIA 
theory.

\subsubsection{KK-monopole/anti-KK-monopole pair}

By uplifting the magnetic KK-dipole solution (\ref{kkdipol}) or 
(\ref{kkdipol2}) to $D=10$, one obtains the following supergravity solution 
for the KK monopole and anti-KK monopole pair:
\begin{eqnarray}
ds^2_{10}&=&-dt^2+dx^2_1+\cdots+dx^2_5
\cr
& &+H^{-1}\left[dx^6+{{2mar\sin^2\theta}\over{r^2-2mr-a^2\cos^2\theta}}
d\varphi\right]^2
\cr
& &+H\left[(r^2-2mr-a^2\cos^2\theta)
\left({{dr^2}\over{r^2-2mr-a^2}}+d\theta^2\right)\right.
\cr
& &\left.+(r^2-2mr-a^2)\sin^2\theta d\varphi^2\right].
\label{10dkk}
\end{eqnarray}

\subsubsection{D6-brane/anti-D6-brane pair}

By first uplifting the KK-monopole/anti-KK-monopole pair solution 
(\ref{10dkk}) to $D=11$ and then compactifying along the $x^6$ 
direction (of the solution in Eq. (\ref{10dkk})), one obtains the 
following solution for the D6-brane and the anti-D6-brane pair:
\begin{eqnarray}
ds^2_{10}&=&H^{-{1\over 2}}\left[-dt^2+(dx^1)^2+\cdots
+(dx^6)^2\right]
\cr
& &+H^{1\over 2}\left[(r^2-2mr-a^2\cos^2\theta)\left({{dr^2}\over
{r^2-2mr-a^2}}+d\theta^2\right)\right.
\cr
& &\left.+(r^2-2mr-a^2)\sin^2\theta d\varphi^2\right],
\cr
\hat{A}_{\varphi}&=&{{2mar\sin^2\theta}\over{r^2-2mr-a^2\cos^2\theta}},
\ \ \ \ \ 
e^{\hat{\Phi}}=H^{-{3\over 2}}.
\label{10dd6}
\end{eqnarray}

\section{Non-marginal Bound States of Fundamental Dipoles}

By applying the $SL(2,{\bf R})$ transformations in section 3 to the 
fundamental dipole solutions presented in section 4, one can construct 
the supergravity solutions for non-marginal bound states of fundamental 
dipoles.  

First, we consider the non-marginal bound states of the (electric or 
magnetic) KK and $H$ dipoles whose $U(1)$ fields are associated with 
different circles.  The Einstein-frame metric $g_{\mu\nu}$ is given by 
Eq. (\ref{fundimet}), and the string-frame metric $g^{\rm str}_{\mu\nu}$ 
is given by Eq. (\ref{strnselec}) for the electric case and Eq. 
(\ref{strnsmag}) for the magnetic case.  We apply the $SL(2,{\bf R})_T$ 
transformation (\ref{sl2rsl2r}) to the fundamental electric KK dipole 
solution (\ref{kkelecdipol}) to construct the supergravity solution for 
the non-marginal bound state of the electric KK and electric $H$ dipoles.  
We use $\omega_T=\left(\matrix{\cos\delta&-\sin\delta\cr\sin\delta&\cos\delta}
\right)$ as the $SO(2)\subset SL(2,{\bf R})_T$ transformation matrix.
The resulting solution has the following form:
\begin{eqnarray}
\eta&=&-\rho={1\over 2}\ln{{r^2-2mr-a^2\cos^2\theta}\over
{r^2-a^2\cos^2\theta}},
\cr
\sigma&=&\ln{{r^2-2mr\cos^2\delta-a^2\cos^2\theta}
\over{\sqrt{(r^2-2mr-a^2\cos^2\theta)(r^2-a^2\cos^2\theta)}}},
\cr
b&=&{{mr\sin 2\delta}\over{r^2-2mr\cos^2\delta-a^2\cos^2\theta}},
\cr
{\cal A}^3_t&=&A^5_t={{2ma\cos\delta\cos\theta}\over{r^2-a^2\cos^2\theta}},
\cr
{\cal A}^1_t&=&B_{4\,t}=-{{2ma\sin\delta\cos\theta}\over
{r^2-a^2\cos^2\theta}}.
\label{eleckkwin}
\end{eqnarray}
The ADM mass $M$ and the electric dipole moments $p^{\rm KK}_{\rm elec}$ 
and $p^{\rm wind}_{\rm elec}$ of this solution are
\begin{equation}
M=m=\sqrt{(m\cos\delta)^2+(m\sin\delta)^2},\ \ \ \  
p^{\rm KK}_{\rm elec}=ma\cos\delta,\ \ \ \ 
p^{\rm wind}_{\rm elec}=-ma\sin\delta.
\label{melkkwindip}
\end{equation}
The expressions for the ADM mass and the dipole moments for the remaining 
cases in the following have the similar forms as Eq. (\ref{melkkwindip}).  
So, we shall not write down them for the remaining cases.  In the case when 
the $U(1)$ fields are magnetic, the solutions are as follows:
\begin{eqnarray}
\eta&=&-\rho=-{1\over 2}\ln{{r^2-2mr-a^2\cos^2\theta}\over
{r^2-a^2\cos^2\theta}},
\cr
\sigma&=&\ln{{r^2-2mr\sin^2\delta-a^2\cos^2\theta}
\over{\sqrt{(r^2-2mr-a^2\cos^2\theta)(r^2-a^2\cos^2\theta)}}},
\cr
b&=&-{{mr\sin 2\delta}\over{r^2-2mr\sin^2\delta-a^2\cos^2\theta}},
\cr
{\cal A}^3_{\varphi}&=&A^5_{\varphi}={{2mar\cos\delta\sin^2\theta}
\over{r^2-2mr-a^2\cos^2\theta}},
\cr
{\cal A}^1_{\varphi}&=&B_{4\,\varphi}=-{{2mar\sin\delta\sin^2\theta}
\over{r^2-2mr-a^2\cos^2\theta}}.
\label{magkkwin}
\end{eqnarray}
The solutions for the case when the KK [winding] $U(1)$ field comes from 
the 4-th [5-th] coordinate are obtained by just applying the ${\bf Z}_2
\subset SL(2,{\bf R})_U$ transformation to the above solutions.  The 
resulting solutions have the opposite sign for $\rho$, and $A^5_{\mu}$ 
and $B_{4\mu}$ respectively replaced by $A^4_{\mu}$ and $B_{5\mu}$.

Similarly, one can apply the $SO(2)\subset SL(2,{\rm R})_U$ transformation 
(\ref{sl2rsl2r}) (with the same transformation matrix as above) to the 
fundamental electric [magnetic] KK or $H$ dipole solution to construct the 
supergravity solution for the non-marginal bound state of two fundamental 
electric [magnetic] KK or $H$ dipoles charged with respect to two KK or 
winding $U(1)$ fields of different circles.  For such case, the real scalar 
$\rho$ undergoes the $SL(2,{\rm R})_U$ transformation and therefore the real 
part $c$ of the complex structure $U$ is induced.  So, the internal metric 
$G_{mn}$ becomes non-diagonal (Cf. Eq.(\ref{scalmet})).  Similarly as above, 
the Einstein-frame metric $g_{\mu\nu}$ is given by Eq. (\ref{fundimet}), and 
the string-frame metric $g^{\rm str}_{\mu\nu}$ is given by Eq. 
(\ref{strnselec}) for the electric case and Eq. (\ref{strnsmag}) for the 
magnetic case.  The solutions have the form similar to the above.  But we 
write down the solutions for the completeness.  For the non-marginal bound 
state of two electric KK dipoles, the solution is given by
\begin{eqnarray}
\eta&=&\sigma={1\over 2}\ln{{r^2-2mr-a^2\cos^2\theta}\over
{r^2-a^2\cos^2\theta}},
\cr
\rho&=&\ln{{r^2-2mr\cos^2\delta-a^2\cos^2\theta}
\over{\sqrt{(r^2-2mr-a^2\cos^2\theta)(r^2-a^2\cos^2\theta)}}},
\cr
c&=&{{mr\sin 2\delta}\over{r^2-2mr\cos^2\delta-a^2\cos^2\theta}},
\cr
{\cal A}^4_t&=&A^4_t={{2ma\cos\delta\cos\theta}\over{r^2-a^2\cos^2\theta}},
\cr
{\cal A}^3_t&=&A^5_t=-{{2ma\sin\delta\cos\theta}\over
{r^2-a^2\cos^2\theta}}.
\label{eleckkkk}
\end{eqnarray}
When the KK $U(1)$ fields are magnetic, the corresponding solution
is the following:
\begin{eqnarray}
\eta&=&\sigma=-{1\over 2}\ln{{r^2-2mr-a^2\cos^2\theta}\over
{r^2-a^2\cos^2\theta}},
\cr
\rho&=&\ln{{r^2-2mr\sin^2\delta-a^2\cos^2\theta}
\over{\sqrt{(r^2-2mr-a^2\cos^2\theta)(r^2-a^2\cos^2\theta)}}},
\cr
c&=&-{{mr\sin 2\delta}\over{r^2-2mr\sin^2\delta-a^2\cos^2\theta}},
\cr
{\cal A}^4_{\varphi}&=&A^4_{\varphi}={{2mar\cos\delta\sin^2\theta}\over
{r^2-2mr-a^2\cos^2\theta}},
\cr
{\cal A}^3_{\varphi}&=&A^5_{\varphi}=-{{2mar\sin\delta\sin^2\theta}\over
{r^2-2mr-a^2\cos^2\theta}}.
\label{magkkkk}
\end{eqnarray}
The non-marginal bound states of two electric (or magnetic) $H$ dipoles 
are related to these configurations through the ${\bf Z}_2\subset 
SL(2,{\bf R})_T$ transformation.  So, the corresponding solution is given by 
Eq. (\ref{eleckkkk}) or (\ref{magkkkk}) with the opposite sign for $\sigma$ 
and with ${\cal A}^4_{\mu}$ and ${\cal A}^3_{\mu}$ respectively replaced by 
${\cal A}^2_{\mu}=B_{5\mu}$ and ${\cal A}^1_{\mu}=B_{4\mu}$.

By applying the $SL(2,{\bf R})_S$ transformation (\ref{sdualtr}) 
to the fundamental dipole solutions, one can construct supergravity solutions 
for the non-marginal bound states of the electric [magnetic] KK dipole and 
the magnetic [electric] $H$ dipole whose $U(1)$ fields are associated with 
different circles.  The following solution for the non-marginal bound state 
of the electric KK dipole and the magnetic $H$ dipole is obtained by 
applying the $SL(2,{\bf R})_S$ transformation to an electric KK dipole 
solution (\ref{kkelecdipol}):
\begin{eqnarray}
\sigma&=&-\rho={1\over 2}\ln{{r^2-2mr-a^2\cos^2\theta}\over
{r^2-a^2\cos^2\theta}},
\cr
\eta&=&\ln{{r^2-2mr\cos^2\delta-a^2\cos^2\theta}
\over{\sqrt{(r^2-2mr-a^2\cos^2\theta)(r^2-a^2\cos^2\theta)}}},
\cr
a&=&{{mr\sin 2\delta}\over{r^2-2mr\cos^2\delta-a^2\cos^2\theta}},
\cr
{\cal A}^3_t&=&A^5_t={{2ma\cos\delta\cos\theta}\over{r^2-a^2\cos^2\theta}},
\cr
{\cal A}^1_{\varphi}&=&B_{4\varphi}=-{{2mar\sin\delta\sin^2\theta}\over
{r^2-2mr-a^2\cos^2\theta}}.
\label{eleckkmagwin}
\end{eqnarray}
In the case where the KK $U(1)$ field $A^5_{\mu}$ is magnetic and the 
winding $U(1)$ field $B_{4\mu}$ is electric, the solution is
\begin{eqnarray}
\sigma&=&-\rho=-{1\over 2}\ln{{r^2-2mr-a^2\cos^2\theta}\over
{r^2-a^2\cos^2\theta}},
\cr
\eta&=&\ln{{r^2-2mr\sin^2\delta-a^2\cos^2\theta}
\over{\sqrt{(r^2-2mr-a^2\cos^2\theta)(r^2-a^2\cos^2\theta)}}},
\cr
a&=&-{{mr\sin 2\delta}\over{r^2-2mr\sin^2\delta-a^2\cos^2\theta}},
\cr
{\cal A}^3_{\varphi}&=&A^5_{\varphi}={{2mar\cos\delta\sin^2\theta}\over
{r^2-2mr-a^2\cos^2\theta}},
\cr
{\cal A}^1_t&=&B_{4t}=-{{2mar\sin\delta\cos\theta}\over
{r^2-a^2\cos^2\theta}}.
\label{magkkelecwin}
\end{eqnarray}
The supergravity solutions for the configurations with non-zero 
${\cal A}^4_{\mu}=A^4_{\mu}$ and ${\cal A}^{2}_{\mu}=B_{5\mu}$ are 
obtained by applying the ${\bf Z}_2\subset SL(2,{\bf R})_U$ transformations 
to the above solutions.  The resulting solutions have the opposite sign for 
$\rho$.  The Einstein-frame metric for all the above cases is given by Eq. 
(\ref{fundimet}).  But since the $D=4$ dilaton $\eta$ has undergone the 
$SL(2,{\bf R})_S$ transformation, the string-frame metric now depends on 
the $SO(2)\subset SL(2,{\bf R})_S$ angle $\delta$.  The string-frame metric 
$g^{\rm str}_{\mu\nu}=e^{\eta}g_{\mu\nu}$ for the case where the KK $U(1)$ 
field is electric and the winding $U(1)$ field is magnetic is
\begin{eqnarray}
g^{\rm str}_{\mu\nu}dx^{\mu}dx^{\nu}&=&-(r^2-2mr\cos^2\delta-
a^2\cos^2\theta)\left[-{{dt^2}\over{r^2-a^2\cos^2\theta}}\right.
\cr
& &\left.+{{dr^2}\over{r^2-2mr-a^2}}+d\theta^2
+{{r^2-2mr-a^2}\over{r^2-2mr-a^2\cos^2\theta}}\sin^2\theta 
d\varphi^2\right],
\label{strmetekmw}
\end{eqnarray}
and for the case where the KK $U(1)$ field is magnetic and the 
winding $U(1)$ field is electric is given by Eq. (\ref{strmetekmw}) with 
$\cos^2\delta$ in the overall factor term replaced by $\sin^2\delta$.

Finally, the supergravity solutions for the non-marginal bound states of 
the $H$ dipole (charged under ${\cal A}^1_{\mu}$ or ${\cal A}^2_{\mu}$) 
and the D dipole (charged under ${\cal B}^a_{\mu}=\hat{B}^{(2)}_{3+a,\mu}$) 
are obtained by applying the $SL(2,{\bf R})$ $S$-duality transformation 
(\ref{sl2rtran}) to the fundamental D dipole solutions (\ref{elecrrdip}) 
and (\ref{magrrdip}).  In this case, the Einstein-frame metric is still 
given by Eq. (\ref{fundimet}).  The solutions are
\begin{eqnarray}
g_{11}&=&{\sqrt{(r^2-2mr-a^2\cos^2\theta)(r^2-2mr\sin^2\delta-a^2\cos^2\theta)}
\over{r^2-a^2\cos^2\theta}},
\cr
g_{ii}&=&\sqrt{{r^2-2mr\sin^2\delta-a^2\cos^2\theta}\over
{r^2-2mr-a^2\cos^2\theta}}\ \ \ \ (i\neq 1),
\cr
e^{\phi}&=&\sqrt{{r^2-2mr\sin^2\delta-a^2\cos^2\theta}\over
{r^2-a^2\cos^2\theta}},\ \ \ \ 
\chi=-{{mr\sin 2\delta}\over{r^2-2mr\sin^2\delta-a^2\cos^2\theta}},
\cr
{\cal B}^1_t&=&{{2ma\cos\delta\cos\theta}\over{r^2-a^2\cos^2\theta}},\ \ \ \ 
{\cal A}^1_t=-{{2ma\sin\delta\cos\theta}\over{r^2-a^2\cos^2\theta}},
\label{elecnsrr}
\end{eqnarray}
for the electric case, and
\begin{eqnarray}
g_{11}&=&{\sqrt{(r^2-a^2\cos^2\theta)(r^2-2mr\cos^2\delta-a^2\cos^2\theta)}
\over{r^2-2mr-a^2\cos^2\theta}},
\cr
g_{ii}&=&\sqrt{{r^2-2mr\cos^2\delta-a^2\cos^2\theta}
\over{r^2-a^2\cos^2\theta}}\ \ \ \ (i\neq 1),
\cr
e^{\phi}&=&\sqrt{{r^2-2mr\cos^2\delta-a^2\cos^2\theta}
\over{r^2-2mr-a^2\cos^2\theta}},\ \ \ \ 
\chi={{mr\sin 2\delta}\over{r^2-2mr\cos^2\delta-a^2\cos^2\theta}},
\cr
{\cal B}^1_{\varphi}&=&{{2mar\cos\delta\sin^2\theta}\over
{r^2-2mr-a^2\cos^2\theta}},\ \ \ \ 
{\cal A}^1_{\varphi}=-{{2mar\sin\delta\sin^2\theta}\over
{r^2-2mr-a^2\cos^2\theta}},
\label{magnsrr}
\end{eqnarray}
for the magnetic case.  The string-frame metric $g^{\rm str}_{\mu\nu}=
e^{\phi}g_{\mu\nu}$ is given by
\begin{eqnarray}
g^{\rm str}_{\mu\nu}dx^{\mu}dx^{\nu}&=&\sqrt{r^2-2mr\sin^2\delta
-a^2\cos^2\theta}\left[-{\sqrt{r^2-2mr-a^2\cos^2\theta}\over
{r^2-a^2\cos^2\theta}}dt^2\right.
\cr
& &+\sqrt{r^2-2mr-a^2\cos^2\theta}\left({{dr^2}\over{r^2-2mr-a^2}}
+d\theta^2\right)
\cr
& &\left.+{{r^2-2mr-a^2}\over{r^2-2mr-a^2\cos^2\theta}}\sin^2\theta 
d\varphi^2\right],
\label{strnsrrelec}
\end{eqnarray}
for the electric case, and
\begin{eqnarray}
g^{\rm str}_{\mu\nu}dx^{\mu}dx^{\nu}&=&\sqrt{r^2-2mr\cos^2\delta
-a^2\cos^2\theta}\left[-{{dt^2}\over\sqrt{r^2-a^2\cos^2\theta}}\right.
\cr
& &+\sqrt{r^2-a^2\cos^2\theta}\left({{dr^2}\over{r^2-2mr-a^2}}
+d\theta^2\right)
\cr
& &\left.+{\sqrt{r^2-a^2\cos^2\theta}\over{r^2-2mr-a^2\cos^2\theta}}
(r^2-2mr-a^2)\sin^2\theta d\varphi^2\right],
\label{strnsrrmag}
\end{eqnarray}
for the magnetic case.

\subsection{Higher-dimensional Embedding}

In this subsection, we oxidize some of the above $D=4$ solutions for the 
non-marginal bound states of $D=4$ fundamental dipoles to obtain the 
delocalized supergravity solutions for the non-marginal bound states of 
the brane/anti-brane pairs.  

\subsubsection{F-string/anti-F-string pair and the pp-wave/anti-pp-wave pair}

By uplifting the solution (\ref{eleckkwin}) for the non-marginal bound 
state of the electric $H$ dipole and the electric KK dipole to $D=10$, one 
obtains the following supergravity solution for the non-marginal bound state 
of F-string/anti-F-string pair and the wave/anti-wave pair:
\begin{eqnarray}
ds^2_{10}&=&-H^{-1}_{\rm F1}H^{-1}_{\rm pp}dt^2
+H^{-1}_{\rm F1}(dx^1)^2+H_{\rm pp}\left[dx^2+{{2ma\cos\delta\cos\theta}\over
{r^2-a^2\cos^2\theta}}dt\right]^2
\cr
& &+(dx^3)^2+\cdots+(dx^6)^2
\cr
& &+(r^2-2mr-a^2\cos^2\theta)\left[{{dr^2}\over{r^2-2mr-a^2}}+d\theta^2\right]
\cr
& &+(r^2-2mr-a^2)\sin^2\theta d\varphi^2,
\cr
\hat{B}^{(1)}_{tx^1}&=&{{2ma\sin\delta\cos\theta}\over
{r^2-2mr\cos^2\delta-a^2\cos^2\theta}},
\ \ \ \ 
\hat{B}^{(1)}_{x^1x^2}={{mr\sin 2\delta}\over{r^2-2mr\cos^2\delta-a^2\cos^2
\theta}},
\cr
e^{\hat{\Phi}}&=&H^{-1}_{\rm F1},
\label{10dfwav}
\end{eqnarray}
where the modified harmonic functions for the F-string/anti-F-string pair 
and the wave/anti-wave pair are respectively
\begin{eqnarray}
H_{\rm F1}&=&{{r^2-2mr\cos^2\delta-a^2\cos^2\theta}\over
{r^2-2mr-a^2\cos^2\theta}},
\cr
H_{\rm pp}&=&{{r^2-a^2\cos^2\theta}\over{r^2-2mr\cos^2\delta-a^2\cos^2\theta}}.
\label{harm10dfwav}
\end{eqnarray}

\subsubsection{D2-brane/anti-D2-brane pair and D0-brane/anti-D0-brane pair}

One can uplift the solution (\ref{10dfwav}) for the non-marginal bound state 
of the F-string/anti-F-string pair and the wave/anti-wave pair to $D=11$ 
and then compactify the longitudinal direction of the $D=11$ pp-wave of 
the resulting solution for the non-marginal bound state of the 
M2-brane/anti-M2-brane pair and the pp-wave/anti-pp-wave pair on $S^1$ 
to obtain the following supergravity solution for the non-marginal bound 
state of the D2-brane/anti-D2-brane pair and the D0-brane/anti-D0-brane pair:
\begin{eqnarray}
ds^2_{10}&=&-H^{-{1\over 2}}_{\rm D0}H^{-{1\over 2}}_{\rm D2}dt^2
+H^{1\over 2}_{\rm D0}H^{-{1\over 2}}_{\rm D2}\left[(dx^1)^2+(dx^2)^2\right]
\cr
& &+H^{1\over 2}_{\rm D0}H^{1\over 2}_{\rm D2}\left[(dx^3)^2+\cdots+(dx^6)^2
\right.
\cr
& &+(r^2-2mr-a^2\cos^2\theta)\left({{dr^2}\over{r^2-2mr-a^2}}+d\theta^2
\right)
\cr
& &\left.+(r^2-2mr-a^2)\sin^2\theta d\varphi^2\right],
\cr
\hat{A}_t&=&{{2ma\cos\delta\cos\theta}\over{r^2-a^2\cos^2\theta}},\ \ \ \ \  
\hat{B}_{tx^1x^2}={{2ma\sin\delta\cos\theta}\over
{r^2-2mr\cos^2\delta-a^2\cos^2\theta}},
\cr
\hat{B}_{x^1x^2}&=&{{mr\sin 2\delta}\over{r^2-2mr\cos^2\delta
-a^2\cos^2\theta}},\ \ \ \ 
e^{\hat{\Phi}}=H^{3\over 2}_{\rm D0}H^{1\over 2}_{\rm D2},
\label{10dd2d0}
\end{eqnarray}
where the modified harmonic functions for the D0-brane/anti-D0-brane pair 
and the D2-brane/anti-D2-brane pair are respectively
\begin{eqnarray}
H_{\rm D0}&=&{{r^2-a^2\cos^2\theta}\over{r^2-2mr\cos^2\delta-a^2\cos^2\theta}},
\cr
H_{\rm D2}&=&{{r^2-2mr\cos^2\delta-a^2\cos^2\theta}\over
{r^2-2mr-a^2\cos^2\theta}}.
\label{harm10dd2d0}
\end{eqnarray}

\subsubsection{Kaluza-Klein dipole and NS5-brane/anti-NS5-brane pair}

The following supergravity solution for the non-marginal bound state of the 
magnetic KK dipole and the NS5-brane/anti-NS5-brane pair can be obtained 
by uplifting the solution (\ref{magkkwin}) for the bound state of the 
magnetic KK dipole and the magnetic $H$ dipole to $D=10$:
\begin{eqnarray}
ds^2_{10}&=&-dt^2+(dx^1)^2+\cdots+(dx^4)^2
\cr
& &+H^{-1}_{\rm KK}\left[dx^5+{{2mar\cos\delta\sin^2\theta}\over
{r^2-2mr-a^2\cos^2\theta}}d\varphi\right]^2
\cr
& &+H_{\rm NS5}\left[(dx^6)^2+H_{\rm KK}\left\{(r^2-2mr-a^2\cos^2\theta)
\right.\right.
\cr
& &\times\left.\left.\left({{dr^2}\over{r^2-2mr-a^2}}+d\theta^2\right)
+(r^2-2mr-a^2)\sin^2\theta d\varphi^2\right\}\right],
\cr
\hat{B}_{\varphi x^6}&=&{{2mar\sin\delta\sin^2\theta}\over
{r^2-2mr\sin^2\delta-a^2\cos^2\theta}},\ \ \ \  
\hat{B}_{x^5x^6}={{mr\sin 2\delta}\over
{r^2-2mr\sin^2\delta-a^2\cos^2\theta}},
\cr
e^{\hat{\Phi}}&=&H_{\rm NS5},
\label{10dkkns}
\end{eqnarray}
where the modified harmonic functions for the magnetic KK dipole and the 
NS5-brane/anti-NS5-brane pair are respectively
\begin{eqnarray}
H_{\rm KK}&=&{{r^2-2mr\sin^2\delta-a^2\cos^2\theta}\over
{r^2-2mr-a^2\cos^2\theta}},
\cr
H_{\rm NS5}&=&{{r^2-a^2\cos^2\theta}\over
{r^2-2mr\sin^2\delta-a^2\cos^2\theta}}.
\label{harm10dkkns}
\end{eqnarray}

\subsubsection{D6-brane/anti-D6-brane pair and D4-brane/anti-D4-brane pair}

One can construct the following supergravity solution for the non-marginal 
bound state of the D6-brane/anti-D6-brane pair and the 
D4-brane/anti-D4-brane pair by first uplifting the solution (\ref{10dkkns}) 
for the KK dipole and the NS5-brane/anti-NS5-brane pair bound state to 
$D=11$ and then compactifying along the $x^5$ direction (of the solution 
in Eq. (\ref{10dkkns})) of the resulting $D=11$ solution for the 
M5-brane/anti-M5-brane pair and the magnetic KK dipole bound state:
\begin{eqnarray}
ds^2_{10}&=&H^{-{1\over 2}}_{\rm D4}H^{-{1\over 2}}_{\rm D6}
\left[-dt^2+(dx^1)^2+\cdots+(dx^4)^2\right]
\cr
& &+H^{1\over 2}_{\rm D4}H^{-{1\over 2}}_{\rm D6}\left[(dx^5)^2+(dx^6)^2\right]
\cr
& &+H^{1\over 2}_{\rm D4}H^{1\over 2}_{\rm D6}\left[
(r^2-2mr-a^2\cos^2\theta)\left({{dr^2}\over{r^2-2mr-a^2}}+d\theta^2\right)
\right.
\cr
& &\left.+(r^2-2mr-a^2)\sin^2\theta d\varphi^2\right],
\cr
\hat{A}_{\varphi}&=&{{2mar\cos\delta\sin^2\theta}\over
{r^2-2mr-a^2\cos^2\theta}},\ \ \ \ \ 
\hat{B}_{\varphi x^5x^6}={{2mar\sin\delta\sin^2\theta}\over
{r^2-2mr\sin^2\delta-a^2\cos^2\theta}},
\cr
\hat{B}_{x^5x^6}&=&{{mr\sin 2\delta}\over
{r^2-2mr\sin^2\delta-a^2\cos^2\theta}},\ \ \ \ \ 
e^{\hat{\Phi}}=H^{-{1\over 2}}_{\rm D4}H^{-{3\over 2}}_{\rm D6},
\label{10dkkd6}
\end{eqnarray}
where the modified harmonic functions for the D4-brane/anti-D4-brane pair 
and the D6-brane/anti-D6-brane pair are
\begin{eqnarray}
H_{\rm D4}&=&{{r^2-a^2\cos^2\theta}\over
{r^2-2mr\sin^2\delta-a^2\cos^2\theta}},
\cr
H_{\rm D6}&=&{{r^2-2mr\sin^2\delta-a^2\cos^2\theta}\over
{r^2-2mr-a^2\cos^2\theta}}.
\label{harm10dkkd6}
\end{eqnarray}

\subsubsection{NS5-brane/anti-NS5-brane pair and pp-wave/anti-pp-wave pair}

The following supergravity solution for the non-marginal bound state of the 
NS5-brane/anti-NS5-brane pair and the pp-wave/anti-pp-wave pair can be 
constructed by uplifting the solution (\ref{eleckkmagwin}) for the magnetic 
$H$ dipole and the electric KK dipole bound state:
\begin{eqnarray}
ds^2_{10}&=&-H^{-1}_{\rm pp}dt^2+(dx^1)^2+\cdots+(dx^5)^2
\cr
& &+H_{\rm NS5}\left[H_{\rm pp}(dx^6+{{2ma\cos\delta\cos\theta}\over
{r^2-a^2\cos^2\theta}}dt)^2\right.
\cr
& &+(r^2-2mr-a^2\cos^2\theta)
\left({{dr^2}\over{r^2-2mr-a^2}}+d\theta^2\right)
\cr
& &\left.+(r^2-2mr-a^2)\sin^2\theta d\varphi^2\right],
\cr
\hat{B}^{(1)}_{\varphi x^i}&=&{{2mar\sin\delta\sin^2\theta}\over
{r^2-2mr-a^2\cos^2\theta}}\ \ \  (i=1,...,5),\ \ \ \ \ 
e^{\hat{\Phi}}=H_{\rm NS5},
\label{10dnspp}
\end{eqnarray}
where the modified harmonic functions for the pp-wave/anti-pp-wave pair 
and the NS5-brane/anti-NS5-brane pair are respectively
\begin{eqnarray}
H_{\rm pp}&=&{{r^2-a^2\cos^2\theta}\over{r^2-2mr\cos^2\delta-a^2\cos^2\theta}},
\cr
H_{\rm NS5}&=&{{r^2-2mr\cos^2\delta-a^2\cos^2\theta}\over
{r^2-2mr-a^2\cos^2\theta}}.
\label{harm10dnspp}
\end{eqnarray}

\subsubsection{NS5-brane/anti-NS5-brane pair and D0-brane/anti-D0-brane pair}

One can uplift the above solution (\ref{10dnspp}) for the 
NS5-brane/anti-NS5-brane pair and the pp-wave/anti-pp-wave pair bound 
state to $D=11$ and then compactify along the longitudinal direction of the 
pp-wave of the resulting $D=11$ solution for the 
M5-brane/anti-M5-brane pair and the pp-wave/anti-pp-wave bound state 
to obtain the following supergravity solution for the non-marginal bound 
state of the NS5-brane/anti-NS5-brane pair and the 
D0-brane/anti-D0-brane pair:
\begin{eqnarray}
ds^2_{10}&=&-H^{-{1\over 2}}_{\rm D0}dt^2+H^{1\over 2}_{\rm D0}
\left[(dx^1)^2+\cdots+(dx^5)^2\right]
\cr
& &+H^{1\over 2}_{\rm D0}H_{\rm NS5}\left[(dx^6)^2+(r^2-2mr-a^2\cos^2\theta)
\left({{dr^2}\over{r^2-2mr-a^2}}+d\theta^2\right)\right.
\cr
& &\left.+(r^2-2mr-a^2)\sin^2\theta d\varphi^2\right],
\cr
\hat{A}_t&=&{{2ma\cos\delta\cos\theta}\over{r^2-a^2\cos^2\theta}},\ \ \ \ \ 
\hat{B}^{(1)}_{\varphi x^i}={{2mar\sin\delta\sin^2\theta}\over{r^2-2mr-
a^2\cos^2\theta}}\ \ \  (i=1,...,5),
\cr
e^{\hat{\Phi}}&=&H^{3\over 2}_{\rm D0}H_{\rm NS5},
\label{10dnsd0}
\end{eqnarray}
where the modified harmonic functions for the D0-brane/anti-D0-brane 
pair and the NS5-brane/anti-NS5-brane pair are respectively
\begin{eqnarray}
H_{\rm D0}&=&{{r^2-a^2\cos^2\theta}\over{r^2-2mr\cos^2\delta-a^2\cos^2\theta}},
\cr
H_{\rm NS5}&=&{{r^2-2mr\cos^2\delta-a^2\cos^2\theta}\over
{r^2-2mr-a^2\cos^2\theta}}.
\label{harm10dnsd0}
\end{eqnarray}

\subsubsection{F-string/anti-F-string pair and the Kaluza-Klein dipole}

The following supergravity solution for the F-string/anti-F-string pair 
and the KK dipole bound state can be constructed by uplifting the solution 
(\ref{magkkelecwin}) for the electric $H$ dipole and the magnetic KK dipole 
bound state to $D=10$:
\begin{eqnarray}
ds^2_{10}&=&-H^{-1}_{\rm F1}dt^2+H^{-1}_{\rm F1}
H^{-1}_{\rm KK}\left(dx^1+{{2mar\cos\delta\sin^2\theta}\over
{r^2-2mr-a^2\cos^2\theta}}d\varphi\right)^2
\cr
& &+(dx^2)^2+\cdots+(dx^6)^2
\cr
& &+H_{\rm KK}\left[(r^2-2mr-a^2\cos^2\theta)
\left({{dr^2}\over{r^2-2mr-a^2}}+d\theta^2\right)\right.
\cr
& &\left.+(r^2-2mr-a^2)\sin^2\theta d\varphi^2\right],
\cr
\hat{B}^{(1)}_{tx^1}&=&{{2mar\sin\delta\cos\theta}\over{r^2-a^2\cos^2\theta}},
\ \ \ \  
e^{\hat{\Phi}}=H^{-1}_{\rm F1},
\label{10dfkk}
\end{eqnarray}
where the modified harmonic functions for the F-string/anti-F-string pair 
and the KK dipole are respectively
\begin{eqnarray}
H_{\rm F1}&=&{{r^2-a^2\cos^2\theta}\over{r^2-2mr\sin^2\delta-a^2\cos^2\theta}},
\cr
H_{\rm KK}&=&{{r^2-2mr\sin^2\delta-a^2\cos^2\theta}\over
{r^2-2mr-a^2\cos^2\theta}}.
\label{harm10dfkk}
\end{eqnarray}

\subsubsection{F-string/anti-F-string pair and D6-brane/anti-D6-brane pair}

The following supergravity solution for the F-string/anti-F-string pair 
and the D6-brane/anti-D6-brane pair bound state can be constructed by 
uplifting the above solution (\ref{10dfkk}) for the F-string/anti-F-string 
pair and the KK dipole bound state to $D=11$ and then compactifying 
along the $x^1$ direction (of the solution in Eq. (\ref{10dfkk})) of 
the resulting $D=11$ solution for the M2-brane/anti-M2-brane pair and 
the KK dipole bound state:
\begin{eqnarray}
ds^2_{10}&=&H^{-{1\over 2}}_{\rm F1}H^{-{1\over 2}}_{\rm D6}
\left[-dt^2+(dx^1)^2\right]+H^{-{1\over 2}}_{\rm D6}\left[(dx^2)^2
+\cdots+(dx^6)^2\right]
\cr
& &+H^{1\over 2}_{\rm D6}\left[(r^2-2mr-a^2\cos^2\theta)\left({{dr^2}
\over{r^2-2mr-a^2}}+d\theta^2\right)\right.
\cr
& &\left.+(r^2-2mr-a^2)\sin^2\theta d\varphi^2\right],
\cr
\hat{A}_{\varphi}&=&{{2mar\cos\delta\sin^2\theta}\over
{r^2-2mr-a^2\cos^2\theta}},\ \ \ \ 
\hat{B}^{(1)}_{tx^1}={{2mar\sin\delta\cos\theta}\over{r^2-a^2\cos^2\theta}},
\cr
e^{\hat{\Phi}}&=&H^{-1}_{\rm F1}H^{-{3\over 2}}_{\rm D6},
\label{10df1d6}
\end{eqnarray}
where the modified harmonic functions for the F-string/anti-F-string pair 
and the D6-brane/anti-D6-brane pair are respectively
\begin{eqnarray}
H_{\rm F1}&=&{{r^2-a^2\cos^2\theta}\over{r^2-2mr\sin^2\delta-
a^2\cos^2\theta}},
\cr
H_{\rm D6}&=&{{r^2-2mr\sin^2\delta-a^2\cos^2\theta}\over
{r^2-2mr-a^2\cos^2\theta}}.
\label{harm10df1d6}
\end{eqnarray}

\subsubsection{F-string/anti-F-string pair and D-string/anti-D-string pair}

The following supergravity solution for the F-string/anti-F-string pair 
and the D-string/anti-D-string pair bound state can be constructed by 
uplifting the solution (\ref{elecnsrr}) for the electric $H$ dipole and 
the electric D dipole bound state to $D=10$:
\begin{eqnarray}
ds^2_{10}&=&H^{-1}_{\rm F1}H^{-{1\over 2}}_{\rm D1}
\left[-dt^2+(dx^1)^2\right]+H^{1\over 2}_{\rm D1}\left[(dx^2)^2+\cdots
+(dx^6)^2\right.
\cr
& &+(r^2-2mr-a^2\cos^2\theta)\left({{dr^2}\over{r^2-2mr-a^2}}
+d\theta^2\right)
\cr
& &\left.+(r^2-2mr-a^2)\sin^2\theta d\varphi^2\right],
\cr
\hat{B}^{(1)}_{tx^1}&=&{{2ma\sin\delta\cos\theta}\over{r^2-a^2\cos^2\theta}},
\ \ \ \ \ 
\hat{B}^{(2)}_{tx^1}=-{{2ma\cos\delta\cos\theta}\over{r^2-a^2\cos^2\theta}},
\cr
e^{\hat{\Phi}}&=&H_{\rm D1}H^{-1}_{\rm F1},\ \ \ 
\hat{\chi}=-{{mr\sin 2\delta}\over{r^2-2mr\sin^2\delta-a^2\cos^2\theta}},
\label{10df1d1}
\end{eqnarray}
where the modified harmonic functions for the F-string/anti-F-string pair 
and the D-string/anti-D-string pair are respectively
\begin{eqnarray}
H_{\rm F1}&=&{{r^2-a^2\cos^2\theta}\over{r^2-2mr\sin^2\delta-a^2\cos^2\theta}},
\cr
H_{\rm D1}&=&{{r^2-2mr\sin^2\delta-a^2\cos^2\theta}\over
{r^2-2mr-a^2\cos^2\theta}}.
\label{harm10df1d1}
\end{eqnarray}

\subsubsection{D5-brane/anti-D5-brane pair and NS5-brane/NS5-brane pair}

The following supergravity solution for the D5-brane/anti-D5-brane pair and 
the NS5-brane/NS5-brane pair bound state can be constructed by uplifting 
the solution (\ref{magnsrr}) for the magnetic D dipole and the magnetic 
$H$ dipole bound state to $D=10$:
\begin{eqnarray}
ds^2_{10}&=&+H^{-{1\over 2}}_{\rm D5}\left[-dt^2+(dx^1)^2+\cdots
+(dx^5)^2\right]
\cr
& &+H_{\rm NS5}H^{1\over 2}_{\rm D5}\left[(dx^6)^2+
(r^2-2mr-a^2\cos^2\theta)\left({{dr^2}\over{r^2-2mr-a^2}}+d\theta^2
\right)\right.
\cr
& &\left.+(r^2-2mr-a^2)\sin^2\theta d\varphi^2\right],
\cr
\hat{B}^{(1)}_{\varphi x^6}&=&{{2mar\sin\delta\sin^2\theta}\over
{r^2-2mr-a^2\cos^2\theta}},\ \ \ \ \ 
\hat{B}^{(2)}_{\varphi x^6}=-{{2mar\cos\delta\sin^2\theta}\over
{r^2-2mr-a^2\cos^2\theta}},
\cr
e^{\hat{\Phi}}&=&H_{\rm NS5}H^{-1}_{\rm D5},\ \ \ \ \ 
\hat{\chi}={{mr\sin 2\delta}\over{r^2-2mr\cos^2\delta-a^2\cos^2\theta}},
\label{10dd5ns5}
\end{eqnarray}
where the modified harmonic functions for the NS5-brane/NS5-brane 
pair and the D5-brane/anti-D5-brane pair are respectively
\begin{eqnarray}
H_{\rm NS5}&=&{{r^2-2mr\cos^2\delta-a^2\cos^2\theta}\over
{r^2-2mr-a^2\cos^2\theta}},
\cr
H_{\rm D5}&=&{{r^2-a^2\cos^2\theta}\over{r^2-2mr\cos^2\delta-a^2\cos^2\theta}}.
\label{harm10dd5ns5}
\end{eqnarray}

\section{Marginal Bound States}

So far we considered the case of non-marginal bound states of the 
brane/anti-brane pairs, in which the harmonic functions of the constituent 
brane/anti-brane pairs are related through the $SO(2)$ rotation angle 
$\delta$ to the modified harmonic function $H$ in Eq. (\ref{fharm}) 
and the overall transverse part of the metric has the form similar 
to that for the ``fundamental'' brane/anti-brane pair solutions.  
In order to gain further insight on more general structure of intersecting 
brane/anti-brane pair solutions, one has to study the case of marginal 
bound states of brane/anti-brane pairs.  Such solutions can be indirectly 
constructed by uplifting the multicharged dipole solutions in $D<10$ 
effective string theories to $D=10$.  However, at this stage the 
construction of such solutions seems to be a difficult task
\footnote{After the first version of this paper appeared in the preprint 
archive, R. Emparan let the author know that such solutions were recently 
constructed by him and his collaborators and satisfy the modified harmonic 
function superposition rules similar to the ones discussed in this paper.}.  
Nonetheless, taking advantage of the fact that the general dilatonic dipole 
solution (\ref{dipsol}) with the special values of the dilaton coupling 
$\alpha=\sqrt{(4-n)/n}$ can be interpreted as the solution for the 
``marginal'' bound state of $n\leq 4$ ``fundamental'' dipoles in $D=4$ 
string theory with the equal dipole moments, one can infer the more general 
cases with non-equal dipole moments from the resulting uplifted $D=10$ 
solutions for marginal bound states of the brane/anti-brane pairs with the 
equal brane dipole moments. 

We consider the (consistently) truncated $D=4$ effective string action 
given in Eq. (\ref{scalact}).  When only $n\leq 4$ of the field strengths 
${\cal F}^i$ are non-zero and the same, the action (\ref{scalact}) can 
be transform to the action (\ref{einmaxdilact}) for the 
Einstein-Maxwell-dilaton system with $\alpha=\sqrt{(4-n)/n}$.  We consider 
the following cases:
\begin{itemize}
\item $\alpha=1$ case: ${\cal F}^2={\cal F}^3\neq 0$ and ${\cal A}^1_{\mu}
={\cal A}^4_{\mu}=0$.  In this case, the real scalars are $\rho=0=\sigma$ 
and $\eta\neq 0$.  To bring the resulting action (\ref{scalact}) to the 
form (\ref{einmaxdilact}) with $\alpha=1$, we define the dilaton $\phi$ 
and the $U(1)$ field strength $F$ in (\ref{einmaxdilact}) as
\begin{equation}
\phi=\eta/2,\ \ \ \  F=\sqrt{2}{\cal F}^2=\sqrt{2}{\cal F}^3.
\label{1u1def}
\end{equation}
So, in terms of the fields in the action (\ref{scalact}), the dipole solution 
is given by
\begin{eqnarray}
g_{\mu\nu}dx^{\mu}dx^{\nu}&=&-{{r^2-2mr-a^2\cos^2\theta}\over
{r^2-a^2\cos^2\theta}}dt^2
\cr
& &+{{(r^2-2mr-a^2\cos^2\theta)(r^2-a^2\cos^2\theta)}
\over{r^2-2mr+m^2\sin^2\theta-a^2\cos^2\theta}}
\left[{{dr^2}\over{r^2-2mr-a^2}}+d\theta^2\right]
\cr
& &+{{r^2-a^2\cos^2\theta}\over{r^2-2mr-a^2\cos^2\theta}}
(r^2-2mr-a^2)\sin^2\theta d\varphi^2,
\cr
\eta&=&\ln{{r^2-2mr-a^2\cos^2\theta}\over{r^2-a^2\cos^2\theta}},\ \ \ \ 
\sigma=\rho=0,
\cr
{\cal A}^2_t&=&{\cal A}^3_t={{2ma\cos\theta}\over{r^2-a^2\cos^2\theta}},
\ \ \ \ 
{\cal A}^1_{\mu}={\cal A}^4_{\mu}=0.
\label{1u1dip}
\end{eqnarray}
The string-frame metric $g^{\rm str}_{\mu\nu}=e^{\eta}g_{\mu\nu}$ is as 
follows:
\begin{eqnarray}
g^{\rm str}_{\mu\nu}dx^{\mu}dx^{\nu}&=&-\left({{r^2-2mr-a^2\cos^2\theta}
\over{r^2-a^2\cos^2\theta}}\right)^2dt^2
\cr
& &+{{(r^2-2mr-a^2\cos^2\theta)^2}
\over{r^2-2mr+m^2\sin^2\theta-a^2\cos^2\theta}}
\left[{{dr^2}\over{r^2-2mr-a^2}}+d\theta^2\right]
\cr
& &+(r^2-2mr-a^2)\sin^2\theta d\varphi^2.
\label{2u1str}
\end{eqnarray}
The ADM mass and the dipole moments of this solution are
\begin{equation}
M=m={1\over 2}m+{1\over 2}m,\ \ \ \ 
p^{\rm elec}_1=p^{\rm elec}_2=ma.
\label{2u1mssdp}
\end{equation}
When uplifted to $D=10$, the solution (\ref{1u1dip}) becomes the following 
supergravity solution for the pp-wave/anti-pp-wave pair along the longitudinal 
direction of the F-string/anti-F-string pair:
\begin{eqnarray}
ds^2_{10}&=&H^{-1}_{\rm F1}\left[-H^{-1}_{\rm pp}+H_{\rm pp}
(dx^1+{{2ma\cos\theta}\over{r^2-a^2\cos^2\theta}})^2\right]+
(dx^2)^2+\cdots+(dx^6)^2
\cr
& &+{{(r^2-2mr-a^2\cos^2\theta)^2}
\over{r^2-2mr+m^2\sin^2\theta-a^2\cos^2\theta}}
\left[{{dr^2}\over{r^2-2mr-a^2}}+d\theta^2\right]
\cr
& &+(r^2-2mr-a^2)\sin^2\theta d\varphi^2,
\cr
\hat{B}_{x^1t}&=&{{2ma\cos\theta}\over{r^2-a^2\cos^2\theta}},
\ \ \ \ \  e^{\hat{\Phi}}=H^{-1}_{\rm F1},
\label{marf1pp}
\end{eqnarray}
where the modified harmonic functions are given by
\begin{equation}
H_{\rm F1}=H_{\rm pp}={{r^2-a^2\cos^2\theta}\over{r^2-2mr-a^2\cos^2\theta}}.
\label{harmmarf1pp}
\end{equation}

\item $\alpha=1/\sqrt{3}$ case: ${\cal F}^2={\cal F}^3=\tilde{\cal F}^1\neq 0$ 
and ${\cal A}^4_{\mu}=0$.  In this case, the real scalars are given by 
$\eta=\sigma=\rho\neq 0$.  So, the fields in the action (\ref{einmaxdilact}) 
are given by
\begin{equation}
\phi={\sqrt{3}\over 2}\eta={\sqrt{3}\over 2}\sigma={\sqrt{3}\over 2}\rho, 
\ \ \ \ 
F=\sqrt{3}{\cal F}^2=\sqrt{3}{\cal F}^3=\sqrt{3}\tilde{\cal F}^1.
\label{3u1def}
\end{equation}
The dipole solution in terms of the field parametrization of the action 
(\ref{scalact}) is therefore given by
\begin{eqnarray}
g_{\mu\nu}dx^{\mu}dx^{\nu}&=&-\left({{r^2-2mr-a^2\cos^2\theta}\over
{r^2-a^2\cos^2\theta}}\right)^{3\over 2}dt^2
\cr
& &+{{[(r^2-2mr-a^2\cos^2\theta)(r^2-a^2\cos^2\theta)]^{3\over 2}}
\over{(r^2-2mr+m^2\sin^2\theta-a^2\cos^2\theta)^2}}\left
[{{dr^2}\over{r^2-2mr-a^2}}+d\theta^2\right]
\cr
& &+\left({{r^2-a^2\cos^2\theta}\over{r^2-2mr-a^2\cos^2\theta}}
\right)^{3\over 2}(r^2-2mr-a^2)\sin^2\theta d\varphi^2,
\cr
\eta&=&\sigma=\rho={1\over 2}\ln{{r^2-2mr-a^2\cos^2\theta}\over
{r^2-a^2\cos^2\theta}},\ \ \ \  \sigma=0,
\cr
{\cal A}^2_t&=&{\cal A}^3_t={{2ma\cos\theta}\over{r^2-a^2\cos^2\theta}},
\ \ 
{\cal A}^1_{\varphi}={{2mar\sin^2\theta}\over{r^2-2mr-a^2\cos^2\theta}}, 
\ \  {\cal A}^4_{\mu}=0.
\label{3u1dip}
\end{eqnarray}
The string-frame metric is given by
\begin{eqnarray}
g^{\rm str}_{\mu\nu}dx^{\mu}dx^{\nu}&=&-\left({{r^2-2mr-a^2\cos^2\theta}
\over{r^2-a^2\cos^2\theta}}\right)^2dt^2
\cr
& &+{{(r^2-2mr-a^2\cos^2\theta)^2(r^2-a^2\cos^2\theta)}
\over{(r^2-2mr+m^2\sin^2\theta-a^2\cos^2\theta)^2}}\left
[{{dr^2}\over{r^2-2mr-a^2}}+d\theta^2\right]
\cr
& &+{{r^2-a^2\cos^2\theta}\over{r^2-2mr-a^2\cos^2\theta}}
(r^2-2mr-a^2)\sin^2\theta d\varphi^2.
\label{3u1str}
\end{eqnarray}
The ADM mass and the dipole moments of the solution are
\begin{equation}
M={3\over 2}m={1\over 2}m+{1\over 2}m+{1\over 2}m,\ \ \ \ 
p^{\rm elec}_1=p^{\rm elec}_2=p^{\rm mag}_1=ma.
\label{3u1mssdp}
\end{equation}
When uplifted to $D=10$, the solution (\ref{3u1dip}) becomes the following 
supergravity solution for the marginal bound state of the 
F-string/anti-F-string pair, the pp-wave/anti-pp-wave pair and the 
NS5-brane/anti-NS5-brane pair:
\begin{eqnarray}
ds^2_{10}&=&H^{-1}_{\rm F1}\left[-H^{-1}_{\rm pp}dt^2
+H_{\rm pp}\left(dx^1+{{2ma\cos\theta}\over{r^2-a^2\cos^2\theta}}dt
\right)^2\right]
\cr
& &+(dx^2)^2+\cdots+(dx^5)^2+H_{\rm NS5}\left[(dx^6)^2\right.
\cr
& &+{{(r^2-2mr-a^2\cos^2\theta)^3}\over
{(r^2-2mr+m^2\sin^2\theta-a^2\cos^2\theta)^2}}
\left({{dr^2}\over{r^2-2mr-a^2}}+d\theta^2\right)
\cr
& &\left.+(r^2-2mr-a^2)\sin^2\theta d\varphi^2\right],
\cr
\hat{B}_{x^1t}&=&{{2ma\cos\theta}\over{r^2-a^2\cos^2\theta}},\ 
\hat{B}_{x^6\varphi}={{2mar\sin^2\theta}\over{r^2-2mr-a^2\cos^2\theta}}, 
\ 
e^{\hat{\Phi}}=H^{-1}_{\rm F1}H_{\rm NS5},
\label{marf1ppns}
\end{eqnarray}
where the modified harmonic functions are given by
\begin{equation}
H_{\rm F1}=H_{\rm pp}=H_{\rm NS5}={{r^2-a^2\cos^2\theta}\over
{r^2-2mr-a^2\cos^2\theta}}.
\label{harmmarf1ppns}
\end{equation}

\item $\alpha=0$ case: ${\cal F}^2={\cal F}^3=\tilde{\cal F}^1=
\tilde{\cal F}^4\neq 0$.  In this case, all the real scalars $\eta$, 
$\rho$ and $\sigma$ are zero.  So, the fields in the action 
(\ref{einmaxdilact}) are given by
\begin{equation}
\phi=0,\ \ \ \ F=2{\cal F}^2=2{\cal F}^3=2\tilde{\cal F}^1=2\tilde{\cal F}^4.
\label{4u1def}
\end{equation}
The dipole solution in terms of the field parametrization of the action 
(\ref{scalact}) is therefore given by
\begin{eqnarray}
g_{\mu\nu}dx^{\mu}dx^{\nu}&=&-\left({{r^2-2mr-a^2\cos^2\theta}\over
{r^2-a^2\cos^2\theta}}\right)^2dt^2
\cr
& &+{{[(r^2-2mr-a^2\cos^2\theta)(r^2-a^2\cos^2\theta)]^2}
\over{(r^2-2mr+m^2\sin^2\theta-a^2\cos^2\theta)^3}}
\left[{{dr^2}\over{r^2-2mr-a^2}}+d\theta^2\right]
\cr
& &+\left({{r^2-a^2\cos^2\theta}\over{r^2-2mr-a^2\cos^2\theta}}
\right)^2(r^2-2mr-a^2)\sin^2\theta d\varphi^2,
\cr
\eta&=&\sigma=\rho=0,
\cr
{\cal A}^2_t&=&{\cal A}^3_t={{2ma\cos\theta}\over{r^2-a^2\cos^2\theta}},
\ \ \ \ 
{\cal A}^1_{\varphi}={\cal A}^4_{\varphi}={{2mar\sin^2\theta}
\over{r^2-2mr-a^2\cos^2\theta}}.
\label{4u1dip}
\end{eqnarray}
Since the $D=4$ dilaton $\eta$ is zero, the string-frame metric is the 
same as the Einstein-frame metric.  The ADM mass and the dipole moments 
of the above solution are
\begin{equation}
M=2m={1\over 2}m+{1\over 2}m+{1\over 2}m+{1\over 2}m,\ \ \ \ 
p^{\rm elec}_1=p^{\rm elec}_2=p^{\rm mag}_1=p^{\rm mag}_4=ma.
\label{4u1mssdp}
\end{equation}
When uplifted to $D=10$, the solution (\ref{4u1dip}) becomes the following 
supergravity solution for the marginal bound state of the 
F-string/anti-F-string pair, the pp-wave/anti-pp-wave pair, the 
NS5-brane/anti-NS5-brane pair and the KK dipole:
\begin{eqnarray}
ds^2_{10}&=&H^{-1}_{\rm F1}\left[-H^{-1}_{\rm pp}dt^2
+H_{\rm pp}\left(dx^1+{{2ma\cos\theta}\over{r^2-a^2\cos\theta}}
\right)^2\right]
\cr
& &+(dx^2)^2+\cdots+(dx^5)^2+H_{\rm NS5}\left[H^{-1}_{\rm KK}
\left(dx^6+{{2mar\sin^2\theta}\over{r^2-2mr-a^2\cos\theta}}d\varphi
\right)^2\right.
\cr
& &+H_{\rm KK}\left\{{{(r^2-2mr-a^2\cos^2\theta)^4}\over
{(r^2-2mr+m^2\sin^2\theta-a^2\cos^2\theta)^3}}
\left({{dr^2}\over{r^2-2mr-a^2}}+d\theta^2\right)\right.
\cr
& &\left.\left.+(r^2-2mr-a^2)\sin^2\theta d\varphi^2\right\}\right],
\cr
\hat{B}_{x^1t}&=&{{2ma\cos\theta}\over{r^2-a^2\cos^2\theta}},\ \ 
\hat{B}_{x^6\varphi}={{2mar\sin^2\theta}\over{r^2-2mr-a^2\cos^2\theta}}, 
\ \ 
e^{\hat{\Phi}}=H^{-1}_{\rm F1}H_{\rm NS5},
\label{marf1ppnskk}
\end{eqnarray}
where the modified harmonic functions are given by
\begin{equation}
H_{\rm F1}=H_{\rm pp}=H_{\rm NS5}=H_{\rm KK}={{r^2-a^2\cos^2\theta}\over
{r^2-2mr-a^2\cos^2\theta}}.
\label{harmmarf1ppnskk}
\end{equation}

\end{itemize}

One can construct other supergravity solutions for the marginal bound 
states of the brane/anti-brane pairs by choosing different combinations 
of non-zero $U(1)$ fields, by using duality transformations in section 3 
or through oxidation to $D=11$ followed by the reduction on $S^1$, just as 
we did in the previous section.  Just as in the non-marginal bound state 
cases studied in the previous section, such solutions will have the form 
similar to the supergravity solutions for the delocalized intersecting 
branes except that the harmonic functions are modified to $H=
{{r^2-a^2\cos^2\theta}\over{r^2-2mr-a^2\cos^2\theta}}$ and the metric 
components in the overall transverse space are modified as
\begin{eqnarray}
& &dr^2+r^2(d\theta^2+\sin^2\theta d\varphi^2)\ \ \ \ \longrightarrow 
\cr
& &{{(r^2-2mr-a^2\cos^2\theta)^n}\over
{(r^2-2mr+m^2\sin^2\theta-a^2\cos^2\theta)^{n-1}}}
\left[{{dr^2}\over{r^2-2mr-a^2}}+d\theta^2\right]
\cr
& &+(r^2-2mr-a^2)\sin^2\theta d\varphi^2,
\label{ovtrmetmod}
\end{eqnarray}
for the case when there are $n\leq 4$ constituents in the marginal bound 
states of brane/anti-brane pairs.  Note, such supergravity solutions 
correspond to the restricted configurations which have only three localized 
overall transverse directions and where all the dipole moments of the 
constituent brane/anti-brane pairs are equal.

As we have seen explicitly from the brane/anti-brane pair (bound state) 
solutions in the previous sections and this section, the supergravity 
solutions for the delocalized brane/anti-brane pairs and their bound 
states (with equal dipole moments for the case of the marginal bound 
states) still satisfy the rules similar to the harmonic function 
superposition rules \cite{tsey} of the (delocalized intersecting) brane 
solutions.  One can summarize the harmonic superposition rules for such 
solutions as follows:  
\begin{itemize}

\item The solutions are still expressed in terms of the ``modified'' 
harmonic functions, each of which is associated with each constituent 
brane/anti-brane pair.  In the case of the 3-dimensional (overall) 
transverse space or with only 3 directions of the (overall) transverse 
space localized, the modified harmonic function is given by
\begin{equation}
H={{r^2-a^2\cos^2\theta}\over{r^2-2mr-a^2\cos^2\theta}}.
\label{modharm}
\end{equation} 
For the non-marginal bound states of two brane/anti-brane pairs with the 
(overall) transverse space with 3 localized directions, the modified 
harmonic functions are
\begin{eqnarray}
H_1&=&{{r^2-a^2\cos^2\theta}\over{r^2-2mr\cos^2\delta-a^2\cos^2\theta}},
\cr
H_2&=&{{r^2-2mr\cos^2\delta-a^2\cos^2\theta}\over
{r^2-2mr-a^2\cos^2\theta}},
\label{modharms}
\end{eqnarray} 
where $\cos^2\delta$ in the modified harmonic functions can be replaced 
by $\sin^2\delta$.

\item The spacetime metric is constructed in terms of the modified 
harmonic functions similarly as the (delocalized intersecting) brane 
solutions.  Namely, the overall worldvolume and the relative transverse 
components of the metric is given by the flat metric times overall 
factors expressed in terms of the products of the ``modified'' harmonic 
functions with the appropriate powers (same as the case of the brane 
solutions).   However, for the (overall) transverse components of the 
metric, although the overall factor is expressed in terms of the products 
of ``modified'' harmonic functions with the same powers as in the case of 
the brane solutions, the (overall) transverse space is no longer 
(conformally) flat.  Namely, the term $(d{\bf y})^2$ in (overall) transverse 
part (with the coordinates ${\bf y}$) of the metric is replaced by a curved 
metric.  In the case of the 3-dimensional (overall) transverse space or 
the (overall) transverse space with only 3 localized directions, the flat 
metric $(d{\bf y})^2$ is replaced in the manner described in Eq. 
(\ref{ovtrmetmod}).  For the case of the non-marginal bound states of two 
brane/anti-brane pairs, the flat metric $(d{\bf y})^2$ is replaced as in Eq. 
(\ref{ovtrmetmod}) with $n=1$.

\item The dilaton is expressed as product of the modified harmonic functions 
with the same powers as the (delocalized intersecting) brane solution cases.

\end{itemize} 

This harmonic function superposition rules can be straightforwardly 
generalized to the case where there is an external magnetic field that 
provides the repulsive force necessary in the force balance between 
brane and anti-brane.  This can be done by using the solution constructed 
in Ref. \cite{emp} in place of the solution (\ref{dipsol}) that we have 
used so far in this paper.  In this case, the generalized harmonic function 
(\ref{modharm}) is modified to
\begin{equation}
H={{r^2-a^2\cos^2\theta}\over A},
\label{magharm}
\end{equation}
where
\begin{eqnarray}
A&\equiv&r^2-2mr-a^2\cos^2\theta+4Bmr\sin^2\theta
\cr
& &+B^2\sin^2\theta\left[(r^2-a^2)^2+a^2(r^2-2mr-a^2)\sin^2\theta\right].
\label{defin}
\end{eqnarray}
Here, $B=2m/(m+a+\sqrt{m^2+a^2})^2$ is the external magnetic field, 
which is tuned in such a way that the conical singularity at $r=r_+$ 
disappears.  Note, when the external magnetic field $B$ is zero, $A=r^2-2mr-
a^2\cos^2\theta$, leading to the previous case with no external 
magnetic field.  And the (overall) transverse space is modified as
\begin{eqnarray}
& &dr^2+r^2(d\theta^2+\sin^2\theta d\varphi^2)\ \ \ \ \longrightarrow 
\cr
& &{{A^n}\over{(r^2-2mr+m^2\sin^2\theta-a^2\cos^2\theta)^{n-1}}}
\left[{{dr^2}\over{r^2-2mr-a^2}}+d\theta^2\right]
\cr
& &+(r^2-2mr-a^2)\sin^2\theta d\varphi^2,
\label{magtransp}
\end{eqnarray}
for the case when there are $n\leq 4$ constituents (with the equal 
dipole moments) in the marginal bound states of brane/anti-brane pairs.

We speculate that even for the full solutions for the (bound states of) 
brane/anti-brane pairs these harmonic superposition rules will hold, with 
the appropriate modified harmonic functions and (overall) transverse 
components of the metric.

\end{document}